\documentclass[conference]{IEEEtran}
\IEEEoverridecommandlockouts

\usepackage{cite}
\usepackage{amsmath,amssymb,amsfonts}
\usepackage{algorithmic}
\usepackage{graphicx}
\usepackage{textcomp}
\usepackage{xcolor}
\usepackage{caption, subcaption}

\def\BibTeX{{\rm B\kern-.05em{\sc i\kern-.025em b}\kern-.08em
    T\kern-.1667em\lower.7ex\hbox{E}\kern-.125emX}}

 \usepackage{caption, subcaption}
 \usepackage{hyperref}
 \usepackage{adjustbox}
 \usepackage{multirow}

 \newcommand{\sr}{\mathbf{r}}
 \newcommand{\lR}{\mathbf{R}}
 
 \newcommand{\bth}{\boldsymbol{\theta}}

\begin{document}
\title{Accelerating Noisy VQE Optimization with Gaussian Processes\\

\thanks{This work was supported by the Office of Science, Office of Advanced Scientific Computing Research Accelerated Research for Quantum Computing Program of the U.S. Department of Energy under Contract No. DE-AC02-05CH11231.}
}

\author{\IEEEauthorblockN{ Juliane M\"uller, Wim Lavrijsen, Costin Iancu, Wibe de Jong}
\IEEEauthorblockA{\textit{Computing Sciences Area} \\
\textit{Lawrence Berkeley National Laboratory}\\
Berkeley, CA, USA \\
$\{$JulianeMueller, WLavrijsen, cciancu,  wadejong$\}$@lbl.gov}
}

\maketitle

\begin{abstract}
Hybrid variational quantum algorithms, which combine a classical optimizer with evaluations on a quantum chip, are the most promising candidates to show quantum advantage on current noisy, intermediate-scale quantum (NISQ) devices.
The classical optimizer is required to perform well in the presence of noise  in the objective function evaluations, or else it becomes the weakest link in the algorithm.
We introduce the use of Gaussian Processes (GP) as surrogate models to reduce the impact of noise and to provide high quality seeds to escape local minima, whether real or noise-induced.
We build this as a framework on top of local optimizations, for which we choose Implicit Filtering (ImFil) in this study.
ImFil is a state-of-the-art, gradient-free method, which in comparative studies has been shown to outperform on noisy VQE problems.
The result is a new method: ``GP+ImFil''.
We show that when noise is present, the GP+ImFil approach finds results closer to the true global minimum in fewer evaluations than standalone ImFil, and that it works particularly well for larger dimensional problems.
Using GP to seed local searches in a multi-modal landscape shows mixed results: although it is capable of improving on ImFil standalone, it does not do so consistently and would only be preferred over other, more exhaustive, multistart methods if resources are constrained.
\end{abstract}


\section{Introduction and Motivation}
Quantum hardware will, for the foreseeable future, consist of devices with a
relatively low number of uncorrected qubits with limited coherence times and connectivity.
The most promising algorithms to exploit the potential of quantum advantage
are therefore those that limit circuit depth and are by design robust against noise.
Hybrid quantum-classical algorithms~\cite{McClean2015,Cerezo2021} such as the Variational Quantum Eigensolver (VQE)~\cite{Peruzzo2013,Tilly2021} and QAOA~\cite{farhi2014quantum,Wang_2018} combine a classical optimizer with evaluations on a quantum chip and fit both these criteria: they allow for control over circuit depth and their iterative nature provides a measure of robustness against noise.
The trade-off is that a large number of evaluations (tens of thousands or
more; a number that increases significantly with noise) on the
quantum chip are necessary for any interesting applications.
By their nature, hybrid quantum-classical algorithms spend time on classical
compute in each iteration cycle.
Classical compute is still easier to scale out than quantum hardware, so techniques that exploit the classical side to reduce the number of needed evaluations on the quantum chip are expected to greatly improve overall performance.
In this paper, we explore one such technique: surrogate model based global optimization coupled with multiple targeted local searches. 

Hybrid variational algorithms minimize the expectation value of an
Hamiltonian $\mathcal{H}$, representing the energy $E$ in the system, as evaluated on the quantum chip.
Finding the minimum energy is  challenging: the energy surface is for most problems a multi-modal landscape, i.e.\ multiple local (and global) optima exist; the analytic description of the energy landscape is usually not accessible, i.e.\ it is black-box; and gradient information is often not available. 
Approximating gradients using finite differences requires many function evaluations, does not scale with increasing problem dimensionality, and  becomes unreliable in the presence of noise.
Finally, local optimization methods, if not provided with a good initial guess, tend to become stuck in a local optimum.

Noise present in the energy function values complicates the optimization problem.
For local optimizers, e.g.\ those based on mesh or stencil refinement, such noise can lead to incorrect sample decisions and  premature convergence onto a noise-induced local minimum. 
``Multistart" methods, which initiate multiple local optimizations from different points, are a typical approach to increase the probability of finding the global minimum on a multi-modal landscape with many local minima (whether real or noise-induced).
However, multistart does not guarantee to always obtain distinct local optima, i.e.\  multiple optimizations may converge to the same solution and it is  unclear how many multistart local optimizations should be carried out.
An efficient global optimizer can help us learn the underlying energy landscape and subsequently seed informed multistart local searches by choosing start points that have good objective function values and are sufficiently disjoint.
Our goals are to alleviate the problems of noise and seeding, by using surrogate models to find interesting regions in the parameter space and then restricting the local search to the close proximity of a diverse set of  starting guesses.  \\

The main contribution of this work is the introduction of a robust algorithm that enables us to efficiently and effectively find good solutions of VQE in the presence of quantum noise.  We use an adaptive Gaussian process model to guide the global optimization search for local regions of attraction from which we start local optimizations on subdomains of the parameter space, thus reducing the a priori requirements on the approximate location of the optimal solution. 
We show  the benefits of our method on solving the Fermi-Hubbard model and compare the efficiency of our approach to the  currently used Implicit Filtering method.

The remainder of this paper is organized as follows. In Section~\ref{sec:related}, we provide a brief review of state-of-the-art optimizers used in quantum hybrid optimization, elaborating on the benefits and pitfalls of various optimization  methods. We describe our simulation of VQE in Section~\ref{sec:vqe} and provide the details about the surrogate model based optimizer in Section~\ref{sec:GPalg}.  Section~\ref{sec:numerics} contains our numerical study on six Hubbard models of different sizes for both noise-free and noisy instances. Finally, Section~\ref{sec:concl} concludes our study and offers future research directions.

\section{Related Work}\label{sec:related}

A wide range of optimization algorithms have been proposed for use with VQE.
In~\cite{Wim} we implemented and applied classical optimizers, including NOMAD~\cite{Nomad}, Implicit Filtering (ImFil)~\cite{ImFil}, SnobFit~\cite{snobfit}, and BOBYQA~\cite{bobyqa} (see scikit-quant~\cite{skquant_web}), and found them to outperform more widely used algorithms, such as those available from SciPy~\cite{scipy_web}.
Simultaneous perturbation stochastic approximation (SPSA~\cite{SPSA}) is a stochastic gradient-free optimization method that is also commonly used because of its application to VQE in Qiskit examples~\cite{qiskitSPSA} and in comparisons with other optimizers~\cite{Kubler}.
In \cite{tutorial}, a comparison between SPSA and ImFil was made and it was shown that ImFil readily outperforms SPSA in the presence of noise, especially at scale.
The authors in~\cite{Kubler} describe a method called individual Coupled Adaptive Number of Shots (iCANS), which uses a stochastic gradient descent method and adaptively decides how many measurements (``shots") of the energy must be used in each iteration and for computing the partial derivatives.
Nakanishi et al.~\cite{Nakanishi} propose a sequential minimal optimization method (NFT) which exploits the special structure present in some parameterized quantum circuits under limiting assumptions and they are able to show superior performance compared to other methods.
However, they do not explicitly test with noise present, and our own tests showed that noise-aware optimizers easily outperform.

In~\cite{Iannelli} and~\cite{shaffer2022}, the authors use Bayesian optimization (BO) and Gaussian process models for VQE optimization tasks. It was shown that this approach outperforms SPSA in terms of convergence with respect to the number of shots used. The authors do not adjust the GP kernel in order to account for the noise in the function values, which can lead to overfitting the model to noise and does not prevent noise-induced local optima.
The authors of~\cite{shiro} propose a stochastic gradient descent (SGD) method for optimizing VQE in which they use a Bayesian optimization approach for adjusting the step size in each iteration of SGD. The authors test their method on different-sized problems and show that it performs better than iCANS, NFT, and the Adam optimizers. Sung et al.~\cite{Sung_2020} use a quadratic model in a trust region to approximate the energy landscape and guide the optimization. This approach performs better than SPSA and BOBYQA.
In~\cite{Learning2Learn}, a classical recurrent neural network is used to provide a good starting guess for a Nelder-Mead optimizer. The authors showed that the number of objective function queries can be significantly reduced as compared to starting Nelder-Mead from a randomly chosen starting point. Here we take a similar approach, but using Gaussian process (GP) surrogate models to guide the initial global search for promising  starting points for local optimizers.

Based on the available literature, combined with our own results and testing, we find that ImFil is an excellent exemplar of the state-of-the-art when noise is present (as is the case on all current hardware).
We will therefore use it as the baseline to compare against\footnote{Code and examples to directly compare ImFil to any other optimizer or method of interest are available from \cite{tutorial}.} and as the local optimizer to improve upon, rather than more well-known, but underperforming, methods such as L-BFGS~\cite{lbfgs} or SPSA, to get a better understanding of the value that surrogate models can provide.

\section{Simulation of VQE}\label{sec:vqe}
The VQE algorithm variationally minimizes the expectation value of an Hamiltonian $\mathcal{H}$, representing the energy $E$ in the system, as evaluated on the quantum chip.
Mathematically, we formulate this optimization problem as:
\begin{equation}
 \min_{\boldsymbol{\theta} \in\Omega}  E(\boldsymbol{\theta}) = \frac{\langle \psi(\bth)|\mathcal{H}|\psi(\bth)\rangle}{\langle \psi(\bth)\psi(\bth)\rangle }, \label{eq:energy}
\end{equation}
where $\Omega \subset \mathbb{R}^d$, $\bth=[\theta_1, \ldots, \theta_d]^T$ and $d$ is the problem dimension.
The representation of $\psi(\bth)$ in terms of $\bth$ is called an {\em Ansatz} and is typically determined by hardware constraints or to ensure symmetry preservation in the problem.

We want to understand in detail under which conditions surrogate models can improve on the classical optimization step of VQE and we will therefore use simulations in order to scan a larger phase space (in particular in scale and depth) than would be possible on currently available hardware devices.
For the simulations, we employ the Hubbard Model (see Section~\ref{ssec:HM}), which is a scientifically relevant problem, provides for multi-modal landscapes with several local minima, and allows easy scaling without changing the underlying fundamentals.
It is therefore very representative of all the typical challenges encountered in VQE problems, providing confidence in the generality of the conclusions from our study.
All simulations are based on Qiskit~\cite{Qiskit}, with noise applied through Qiskit's Aer simulator as applicable.

\subsection{Noise-free simulation}
VQE is a NISQ-era algorithm that will be superseded by the quantum phase estimation~\cite{nielsen00} algorithm once fully error corrected quantum computing for circuits of sufficient depth has been achieved.
Nevertheless, it is useful to consider the noise-free case as a reference point.

Two sources of errors remain even in the absence of noise: sampling error, due to the nature of quantum measurement; and approximation error, an artefact of having to map the unitary matrix that describes the science problem onto a circuit of gates.
The former we ignore in this case, as it can be made as small as required to achieve the desired precision with relatively low resource costs: although sampling error improves only with $\sqrt{N}$, with $N$ the number of samples, single shot wall clock time is very low compared to other latencies in the system.
The same is not true for the latter: a circuit can approximate a unitary to arbitrary precision, but at a worst case cost of exponential scaling.
We will therefore simulate actual mapped circuits, as opposed to unitary matrices, in Qiskit.

We opt to directly calculate the expectation value $\langle \psi(\bth)|\mathcal{H}|\psi(\bth)\rangle$ from the final state to obtain the estimated energy $E$, rather than simulate the necessary partial tomography, since this is mathematically equivalent to measuring the $\mathcal{H}$ components and summing them, when sampling and measurement errors are not considered.

\subsection{Noisy simulation}

\begin{figure}[h]
  \centering
    \includegraphics[scale=.4]{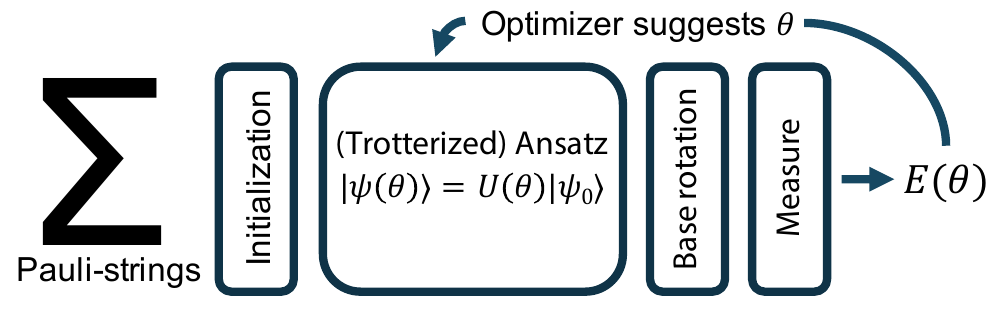}
  \caption{Structure of the VQE algorithm}\label{fig:vqe_structure}
\end{figure}

The impact of noise differs depending on which step in the VQE algorithm (see Fig.~\ref{fig:vqe_structure}) it originates.
Errors in state initialization and final base rotations are rare (these are all single qubit gates) and can be filtered out due to their outsize impact.
State preparation errors are due to drive errors in the portion of the circuit that represents the Ansatz.
These lead to an upward bias in the results, both in the mean and in individual experiment results: any prepared state that is not the global minimum will result in a higher energy estimate.
Due to the nature of quantum mechanics, any experimental result is always calculated from a large set of samples.
This means that drive errors increase the minimum step size in optimization parameters to result in a statistically significant difference in outcome, in effect ``blurring" the optimization surface.
Measurement errors due to misclassification bias the energy estimate as well, but only in the average and the bias direction depends on the specific Hamiltonian considered: individual experiment results can both over- and undershoot the actual minimum energy value.

Sampling errors are bias-free and thus have the least impact of all noise sources.
We do include them in our studies, but measurement errors are what increases the chances of the optimizer getting trapped in a noise-induced local minimum and are thus the most important to consider.

\subsection{The Fermi-Hubbard Model}\label{ssec:HM}

\begin{figure}[h]
  \vspace{-0.2in}
  \centering
    \includegraphics[scale=.4]{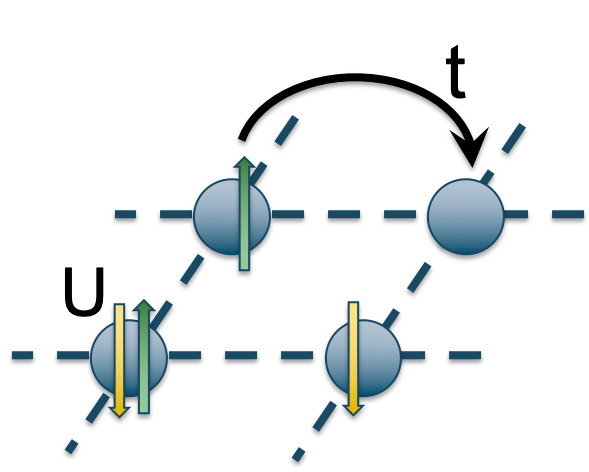}
  \caption{Basic Fermi-Hubbard Model on a 2x2 grid. $U$ denotes  the interaction term and $t$ is the hopping term.}\label{fig:basic_hubbard}
\end{figure}

We will employ the Fermi-Hubbard Model~\cite{Hubbard1963} (HM), used in fields such as solid-state physics to explain phase transitions and other correlated behaviors, as an exemplar of a realistic scientific problem where quantum computers are expected to have an advantage over classical devices.
There are several variations of the HM, but in essence, the model describes interacting particles on a lattice, see Fig.~\ref{fig:basic_hubbard}. Despite its apparent simplicity, the known algorithms to solve the HM scale exponentially on a classical computer, unless further approximations are made.

With reference to Fig.~\ref{fig:basic_hubbard}, the model that we consider forms a periodic grid that is varied to scale the problem size.
The grid is populated with electrons, which determines overall symmetry and allows a (limited) control on the total number of optimization parameters $\bth$.
By their nature, two electrons cannot occupy the same quantum state, thus there can be at most two per site (one spin-up, one spin-down).
The model is characterized by an interaction term $U$, which only contributes if a spin-up and a spin-down electron occupy the same site; and by a ``hopping" term $t$, representing the kinetic energy in the system.
In this basic model, only hops to neighboring sites are considered.

We use the generalized unitary coupled cluster Ansatz for singles and doubles (UCCSD), as it best matches the symmetries in the model.
UCCSD results in long circuits, but we do not optimize them to guarantee consistent interpretation of the results when noise is added.
In a practical, experimental setting, (noise-aware) circuit optimization\cite{bqskit_web} should be used and UCCSD responds well if the encoding is not fully global, e.g.\ by using Bravyi-Kitaev.\cite{Bravyi_2002}

\section{Surrogate Model Based Optimization}\label{sec:GPalg}
The VQE problem~\ref{eq:energy} is a computationally expensive black-box optimization problem since the energy is evaluated with a simulation (on the quantum processor or in software) and we do not have an analytic description of the objective function.
Gradient information is unavailable, and evaluating the objective function is time consuming. For such problems, the goal is to query the objective function as few times as possible during the search for optimal parameters. To this end, surrogate models have been widely used throughout the literature (see for example~\cite{Forrester, somi,mopaper, Jones1998}). Our surrogate model $s(\bth)$ serves as a computationally cheap approximation of the energy objective function: $E(\bth) = s(\bth) +e(\bth)$, where $e(\bth)$ denotes the difference between the two. The surrogate model then  guides the iterative sampling decisions during the search for the optimal solution. 

\subsection{General Surrogate Model Optimization Algorithms}
Surrogate model guided optimization algorithms generally follow the same steps. 
First, an initial experimental design $\mathcal{P}=\{\bth_1, \bth_2, \ldots, \bth_{n_0}\}$ is created, e.g.\ by using  Latin hypercube sampling~\cite{Ye2000}. The initial design can also be augmented with points in the search space that are known to have good performance. 
The energy objective function $E$ is evaluated at all design points.
Based on the input-output data pairs $\{(\bth_i, E(\bth_i) )\}_{i=1}^{n_0}$, a surrogate model is constructed. Generally, different surrogate models can be used, including radial basis functions~\cite{Powell1992}, Gaussian process (GP) models~\cite{Matheron1963, Jones1998}, polynomial regression models, etc. The surrogate models are comparatively cheap to build and evaluate, and thus an efficient option for guiding the optimization search.

An auxiliary optimization problem is formulated and solved over the surrogate surface to decide which point(s) in the parameter space to evaluate with the expensive objective function.
The evaluations at the selected point(s) and the new input-output data pair(s) are then used to update the surrogate model and the process iterates until a stopping criterion (e.g.\ a maximum number $B_{GP}$ of objective function evaluations) has been reached. This process is also referred to as active learning in the literature.

We employ GP models as surrogates because they can be used to approximate noisy data and provide an uncertainty estimate together with the objective function value predictions.
One disadvantage of GPs can be their computational overhead at scale.
Each time the GP is updated, an optimization sub-problem must be solved in search of the optimal GP kernel hyperparameters, which quickly becomes computationally demanding as the number of training samples grows.
This is not of immediate concern for current hardware, especially since the number of allowable function evaluations is limited (e.g.\ due the quantum processor being a shared resource, or because of practical issues such as calibration drift).
This may change, however, with improved hardware and reduced latencies, but wall-clock performance of GPs has already been improved for use with large data sets through GPU acceleration~\cite{gpytorch}.

\subsection{Gaussian Process Models}

In GP modeling~\cite{Jones1998, Jones2001, Mockus}, we  assume that the expensive function is the realization of a stochastic process and we write the GP surrogate model $s_{\text{GP}}$ as:
\begin{equation}
s_{\text{GP}}(\bth) =\mu +Z(\bth),    
\end{equation}
where $\mu$ is the mean of the stochastic process, and $Z(\bth) \sim \mathcal{N}(0,\sigma^2)$  represents the deviation from the mean.  
Assume we have sampled at $k$ points in the parameter space $\Omega$, and  have obtained the data pairs $\{(\bth_i, E(\bth_i))\}_{i=1}^k$. 
The GP prediction at a new point $\bth_{\text{new}}$ is  the realization of a random variable that is distributed as  $\mathcal{N}(\mu, \sigma^2)$. The correlation between two random variables $Z(\bth_m)$ and $Z(\bth_l)$ depends on the chosen kernel. For example, the correlation for a squared exponential kernel  is defined as  
\begin{equation}\label{eq:corr}
    K_{\text{SE}}(Z(\bth_m), Z(\bth_l)) = \exp\left(-\sum_{j=1}^d \tau_j |\theta_m^{(j)} -\theta_l^{(j)}|^2\right),
\end{equation}
where the length scales $\tau_j$ determine how quickly two points become uncorrelated in the $j$th dimension and  $\theta_m^{(j)}$ and $\theta_l^{(j)}$ denote the $j$th component of the vectors $\bth_m$ and $\bth_l$, respectively. Maximum  likelihood estimation is used to determine parameters $\mu$, $\sigma^2$, $\tau_j$, and the GP prediction at a new point $\bth_{\text{new}}$ is 
\begin{equation}
    s_{\text{GP}}(\bth_{\text{new}}) = \hat{\mu} +\sr^T\lR^{-1}(\mathbf{e}-\mathbf{1}\hat{\mu}),
\end{equation}
where the $(m,l)$th element of the $(k\times k)$ covariance matrix  $\lR$ is given by~(\ref{eq:corr}), $\mathbf{e} =[E(\bth_1), \ldots, E(\bth_k)]^T$,  $\mathbf{1}$ is a vector of 1s of appropriate dimension,  $\sr=[K_\text{SE}(Z(\bth_{\text{new}}), Z(\bth_1)), \ldots, K_\text{SE}(Z(\bth_{\text{new}}), Z(\bth_k))]^T$,  
\begin{equation}
    \hat{\mu} = \frac{\mathbf{1}^T\lR^{-1}\mathbf{e}}{\mathbf{1}^T\lR^{-1}\mathbf{1}} \quad \text{ and } \quad
    \hat\sigma^2 = \frac{(\mathbf{e}-\mathbf{1}\hat{\mu})^T\lR^{-1}(\mathbf{e}-\mathbf{1}\hat{\mu})}{k},
\end{equation}
and thus the corresponding mean squared error follows as:
\begin{equation}
    \epsilon^2(\bth_{\text{new}}) = \hat\sigma^2 \left( 1-\sr^T\lR^{-1}\sr +\frac{(1-\mathbf{1}^T\lR^{-1}\sr)^2}{\mathbf{1}^T\lR^{-1}\mathbf{1}}   \right).
\end{equation}

When selecting a new sample point in each iteration, we  use the expected improvement criterion (see~\cite{Jones1998}):
\begin{equation}\label{eq:EI}
    \mathbb{E} I(\bth) =\epsilon(\bth) (g \Phi(g)+\phi(g)), \text{ with } g=\frac{E^{\text{best}}-s_{\text{GP}}(\bth)}{\epsilon(\bth)},
\end{equation}
where $E^{\text{best}}$ is the best energy function value found so far, $\epsilon(\bth)=\sqrt{\epsilon^2(\bth)}$, and $\Phi$ and $\phi$ are the normal cumulative distribution and density function, respectively. The expected improvement function is zero at points where $E$ has already been evaluated and positive everywhere else. (\ref{eq:EI}) is   maximized over all $\bth \in \Omega$ in order to select a new sample point $\bth_{\text{new}}$. One drawback of this approach is that the expected improvement function~(\ref{eq:EI}) is multi-modal, and thus a global optimization algorithm is needed to find various local maxima. However, even with a global optimizer, we cannot  guarantee that the newly chosen point $\bth_{\text{new}}$ will be a global maximum.

\begin{figure}[htbp]
     \centering
         \includegraphics[width=\linewidth]{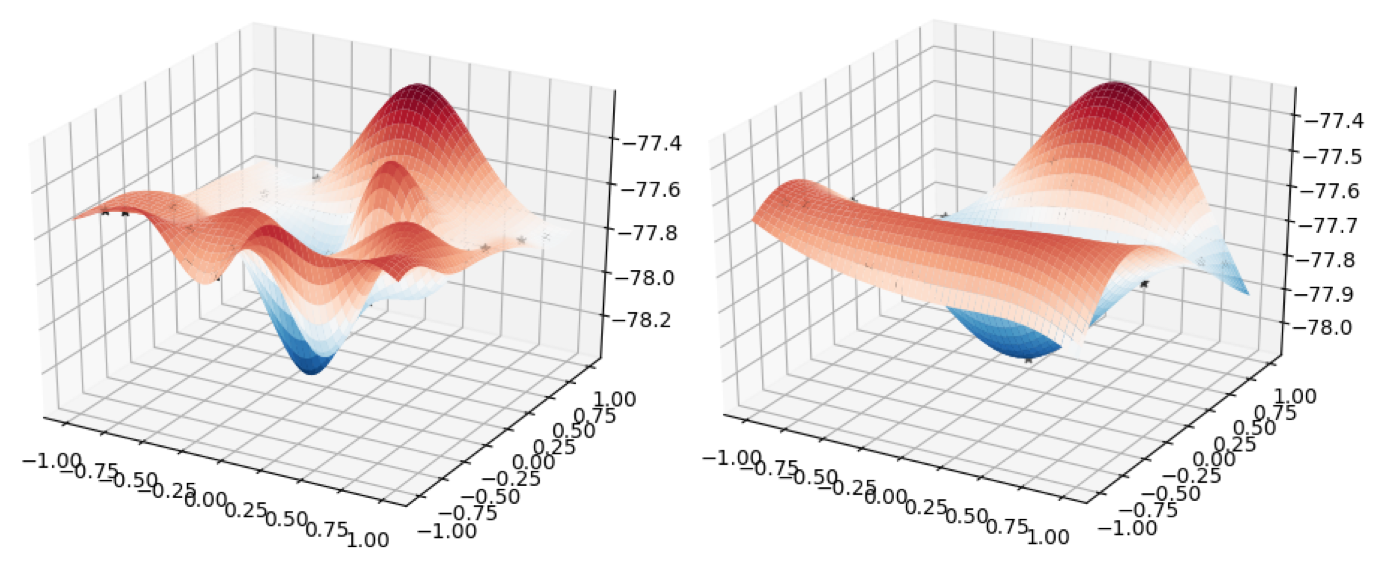}
        \caption{GP with squared exponential kernel fit to noisy  data  ($\sigma_{\text{noise}}=0.05$) without (left)  and with  (right) the addition of the white noise kernel.}
        \label{fig:GP_noisy}
\end{figure}

The squared exponential kernel in~(\ref{eq:corr}) is widely used, in particular for functions that do not contain noise, and the resulting GP model will interpolate the training function values. However, when noise is present, an interpolating model will overfit the function values and interpolate the noise, which may lead to rugged surfaces with many noise-induced local minima (see Fig.~\ref{fig:GP_noisy}, left, for an illustration). When dealing with noisy function values, we  add a white noise kernel to the squared exponential kernel: 
\begin{equation}
    K_{\text{WN}}(Z(\bth_m), Z(\bth_l)) = 
    \begin{cases} \sigma_{\text{noise}}, 
    &\text{ if } \bth_m = \bth_l\\
    0, &\text{ else}
    \end{cases},
\end{equation}
which allows us to estimate the noise level in the data.
The maximum likelihood also estimates $\sigma_{\text{noise}}$. The addition of the white noise kernel prevents the GP from being interpolating, enabling us to capture the underlying  global trends  of the function we are approximating and thus making optimization easier (see Fig.~\ref{fig:GP_noisy}, right).   Note that for noise-free simulations, the addition of the white kernel does not deteriorate the GP approximation significantly.


\subsection{From Global to Local Search}\label{ssec:globloc}
Many optimizers require an initial starting guess and then perform a local optimization from or in the region of the initial.
There is then a strong dependence of the algorithm's performance on the provided starting point.
The goal of the present work is to use the  GP model based optimizer  to identify multiple suitable starting points for local searches.  Note that for this local search, any suitable method can be used. We use ImFil as implemented in scikit-quant~\cite{skquant_web} as justified in Section~\ref{sec:related}.
ImFil is a steepest descent method targeted at noisy optimization problems that have  bound constraints. It uses coordinate search and approximates gradients to inform the iterative sampling decisions.

The specific steps of our algorithm are provided below:
\begin{enumerate}
    \item  \textit{Initialization:} Create a large space filling design, $\mathcal{P} = \{\bth_1, \ldots, \bth_{n_0}  \}$\ and evaluate the objective function at all points in $\mathcal{P}$: $\mathcal{F}=\{E(\bth_1), \ldots, E(\bth_{n_0})\}$; Set the maximum allowable function evaluations for the GP model search as $B_{\text{GP}}$ and for the local search iterations as $B_{\text{loc}}$;  Set the maximum number of allowed local searches as $B_{\text{start}}$; Set the weight pattern $\mathcal{W} = \{w_1, \ldots, w_{B_{\text{start}}-1}\}$; Choose a kernel function for the GP.
    \item  Set $k\leftarrow n_0$.\
    \item \textit{GP iteration:} While $k<B_{\text{GP}}$:
    \begin{enumerate}
        \item \textit{Build the GP}: Use all input-output pairs $\{(\bth_i, E(\bth_i))\}_{i=1}^{k}$ to build the GP.
        \item \textit{Sample point selection}: Maximize the expected improvement function~(\ref{eq:EI}) to select a new evaluation point $\bth_{\text{new}}$.
        \item Evaluate $E_{\text{new}}=E(\bth_{\text{new}})$.
        \item $\mathcal{P} \leftarrow \mathcal{P}\cup\{\bth_{\text{new}}\}$,   $\mathcal{F} \leftarrow \mathcal{F}\cup\{E_{\text{new}}\}$,    $k \leftarrow k+1$.
    \end{enumerate}
    \item \textit{Starting point initialization}: Select the best evaluated point found so far, $\bth^{\text{best}}$ and set $\mathcal{X}_{\text{start}} =\{ \bth^{\text{best}}$\}.
    \item \textit{Scale function values to [0,1]}:  $V_E(\bth_i) = (E(\bth_i)-E^\text{min})/(E^\text{max}-E^\text{min}) \ \forall \bth_i\in\mathcal{P}$, where $E^\text{max}$ and $E^\text{min}$ are the largest and smallest function values in $\mathcal{F}$.
    \item Set $n_s\leftarrow1$.
    \item \textit{Starting point selection iteration:} While $n_s<B_{\text{start}}$:
    \begin{enumerate}
        \item \textit{Compute distances:} $\Delta_i = \Delta(\bth_i, \mathcal{X}_{\text{start}}) = \min_{\bth \in \mathcal{X}_{\text{start}}} \|\bth_i -\bth\|_2 \ \forall \bth_i \in \mathcal{P}\setminus  \mathcal{X}_{\text{start}}$.
        \item \textit{Scale distances to [0,1]}: $V_\Delta(\bth_i) = (\Delta^\text{max}-\Delta_i)/(\Delta^\text{max}-\Delta^\text{min}) \ \forall \bth_i \in\mathcal{P}\setminus  \mathcal{X}_{\text{start}}, $ where $\Delta^\text{max}$ and $\Delta^{\text{min}}$ are the largest and smallest distances.
        \item \textit{Compute weighted scores:} $V_\text{tot}(\bth_i) =w_{n_s} V_E(\bth_i) +(1-w_{n_s})V_\Delta(\bth_i)$.
        \item Select $\bth_{\text{add}} \in\arg \min \{V_\text{tot}(\bth_i), \bth_i \in\mathcal{P}\setminus  \mathcal{X}_{\text{start}}\}$.
        \item Set $\mathcal{X}_{\text{start}} \leftarrow \mathcal{X}_{\text{start}} \cup \{\bth_{\text{add}}\}$, $n_s \leftarrow n_s+1$. 
    \end{enumerate}
\item   Set $m\leftarrow 0$; ctr$\leftarrow 0$.
\item \textit{Local search iteration:} While $m<B_\text{loc}$ and $ \text{ctr}<B_{\text{start}}$:
\begin{enumerate}
    \item  Set $\bth_\text{start} \leftarrow \mathcal{X}_\text{start}[\text{ctr}]$.
    \item Perform  ImFil search  from $\bth_\text{start}$ in reduced bounding box defined by $\bth_\text{start}^{(j)} \pm 0.05$ ; denote the acquired sample points by $\mathcal{P}_\text{loc}$, $n_\text{eval} \leftarrow|\mathcal{P}_\text{loc}|$ and the corresponding function values by $\mathcal{F}_\text{loc}$.
    \item $\mathcal{P} \leftarrow\mathcal{P} \cup \mathcal{P}_\text{loc}$; $\mathcal{F} \leftarrow\mathcal{F} \cup   \mathcal{F}_\text{loc}$;  $m\leftarrow m+n_\text{eval}$; ctr$\leftarrow$ ctr+1.
\end{enumerate}
\item Return the best solution found during optimization.
\end{enumerate}

Here,  $B_\text{GP}$ defines the  budget of expensive function evaluations that are allocated to the optimization with the GP.  $B_\text{loc}$ defines the total number of function evaluations allowed during the local search. Since a single local search does not necessarily use up all of $B_\text{loc}$, we define a maximum number of allowed local searches $B_\text{start}$. 

After the GP iterations have finished, we select $B_\text{start}$ different points from the sample set $\mathcal{P}$ that we will use as seeds for the local search. The goal is to select these points such that (1)  they have good (low) predicted function values and (2)  they are sufficiently spacely separated to minimize the risk of the local search  ending up in the same minimum. The best point found during the GP iterations is used to initialize the set of stating points $\mathcal{X}_\text{start}$. Then,  we iteratively add a new point from  $\mathcal{P}\setminus  \mathcal{X}_{\text{start}}$ to $\mathcal{X}_\text{start}$ by using a score that trades off the criteria (1) and (2). To this end,  we  define a weight pattern $\mathcal{W}$ with elements $w_j\in[0,1)$ which helps us to balance both criteria.

The final step are the multistart local searches. To focus the local search on the vicinity of the starting guess, the search space  defined by upper and lower bounds is reduced.
After each local search, the sets containing all sample points $\mathcal{P}$ and function values $\mathcal{F}$ are updated with the points in $\mathcal{P}_\text{loc}$ and function values in $\mathcal{F}_\text{loc}$, respectively,  that were obtained during the local search.
New local searches are started until either  the budget $B_\text{loc}$  or the maximum number of local searches has been reached.
Eventually, the point corresponding to the lowest function value   is returned as solution. 

There are three places in the algorithm where parallelization can be exploited: (1) in the evaluation of the initial experimental design, all $n_0$ points can be evaluated simultaneously; (2) during the GP iterations, multiple points can be selected for evaluation in each iteration by using different local maxima of the expected improvement function as new sample points; and (3) the $B_{\text{start}}$ local searches can be executed in parallel. 

There are several hyperparameters that can be adjusted to improve the performance of the algorithm.
These include the size of the initial experimental design ($n_0$), the maximum number of GP iterations $B_\text{GP}$, the number of local searches $B_\text{start}$ and evaluations allocated to the local search $B_\text{loc}$, the weight pattern $\mathcal{W}$,  the kernel choice for the GP, and the size for the local search box for each ImFil run after the GP iterations.

There is a direct trade-off between $n_0$ and $B_\text{GP}$. The larger $n_0$, the better the initial GP model. However, this also means that fewer points will be chosen during the GP iteration,  which then has fewer iterations to adapt to and hone in on promising regions. Similarly, the total number of allowed function evaluations   must be separated into $B_\text{GP}$ and $B_\text{loc}$. There is generally no clear guidance for when to stop the GP iteration. Besides setting the upper limit   $B_\text{GP}$, one can stop the GP iterations after a certain number of consecutively failed improvement trials. However, there is no guarantee that the GP did not get stuck in a local optimum if this happens, especially not in high dimensional spaces. The number of local searches that can be afforded depends on  $B_\text{loc}$ and how many evaluations are needed by each local search, which will depend on the starting guess.


\section{Numerical Study}\label{sec:numerics}
We compare our proposed algorithm (``GP+ImFil'') to using ImFil standalone (``ImFil''), and to a GP-only method (``GP'') where we do not use a local search, but rather spend the full budget on GP iterations (i.e.\ Bayesian optimization).
Since the optimization algorithm contains stochasticity (there is randomness in the initial experimental design and in solving expected improvement problems for the GP methods), we perform three runs of each algorithm for each test problem to get an idea of the variability of the results. We limit each algorithm to 1000 function evaluations. For the methods that use the GP, we use an initial experimental design with $2(d+1)$ points and, for the GP+ImFil method, we perform an additional  $8(d+1)$ evaluations during the GP iterations before ImFil starts. In order to obtain a fair comparison with using ImFil only, we generate the same initial experimental design for ImFil, and then use the points in the initial design to seed the multiple restarts of the method. 


\subsection{Test Problem Setup}
\begin{figure}[htbp]
     \centering
     \begin{subfigure}[b]{0.45
     \textwidth}
         \centering
         \includegraphics[width=\textwidth]{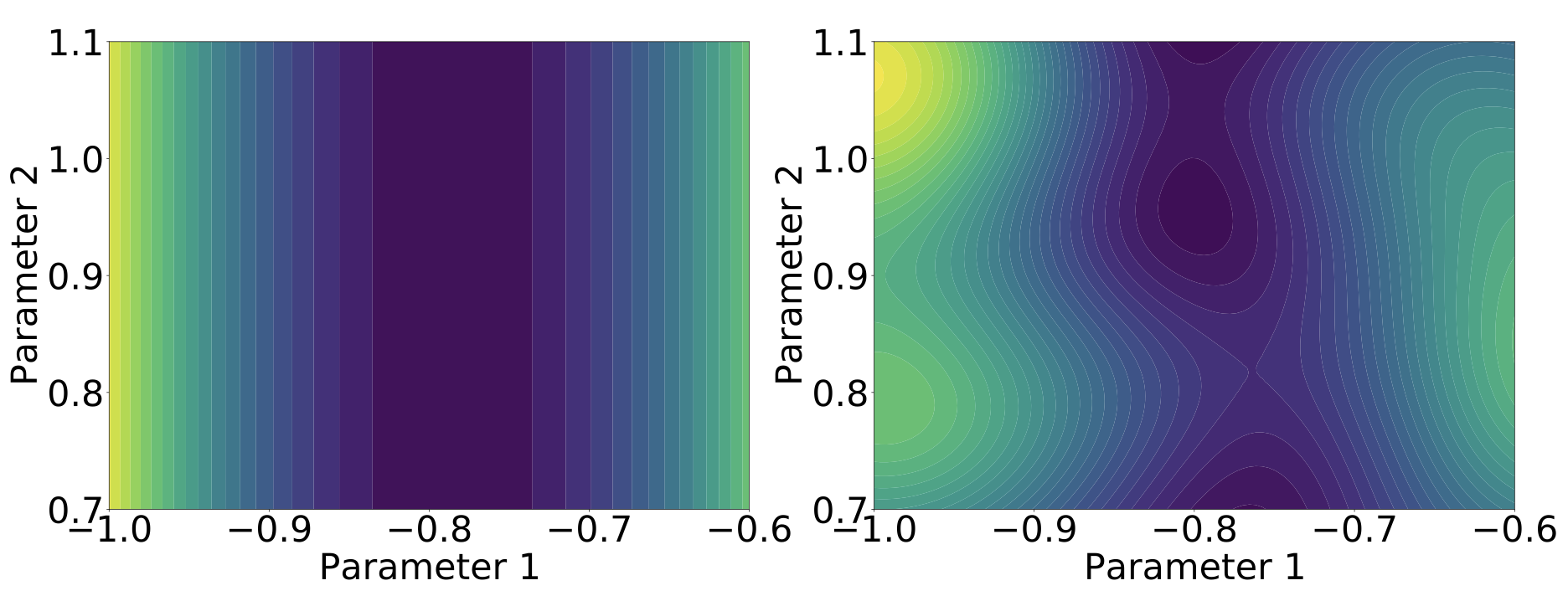}
         \caption{Energy landscape of the noise-free (left) and noisy (right) Hubbard model with 2x1 grid and (1,0) filling.}
     \end{subfigure}
     \hfill
     \begin{subfigure}[b]{0.45\textwidth}
         \centering
         \includegraphics[width=\textwidth]{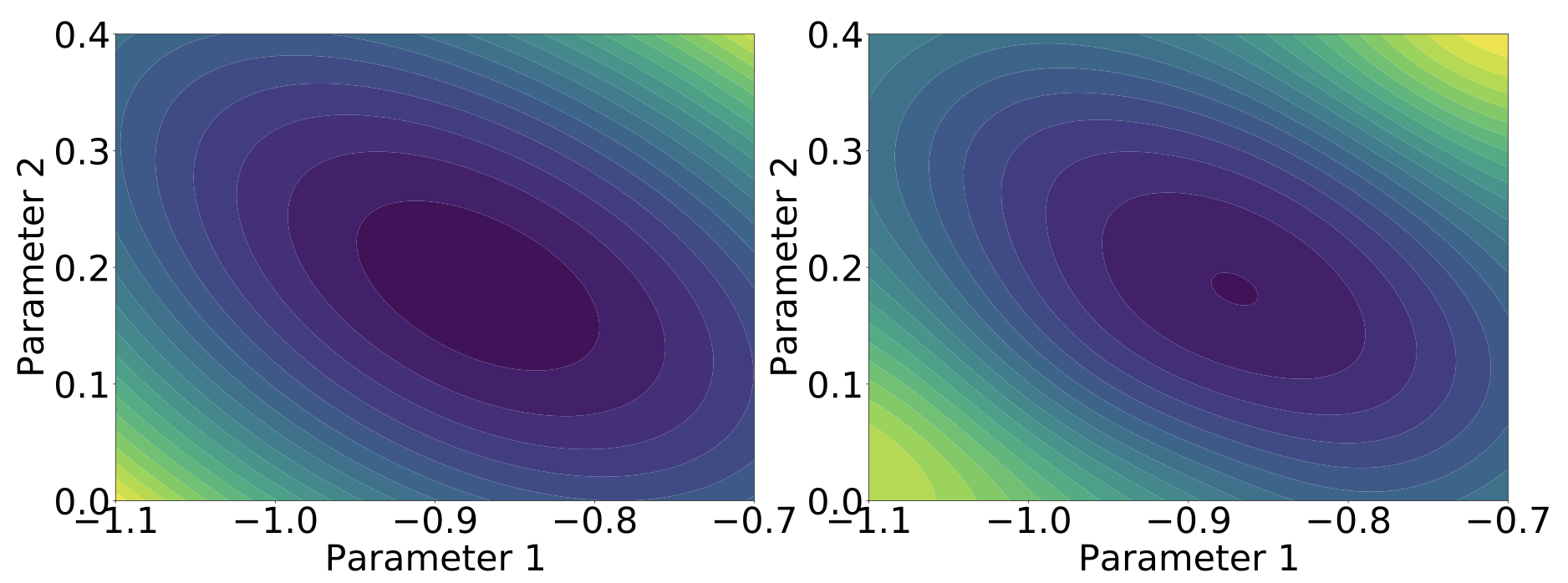}
         \caption{Energy landscape of the noise-free (left) and noisy (right) Hubbard model with 2x1 grid and (1,1) filling.}
     \end{subfigure}
        \caption{Two basic examples (a 2x1 grid, which is described by 2 parameters) of energy landscapes of a noise-free and noisy Hubbard model. We can clearly see how the noise impacts the shape of the energy landscapes and creates local optima.}
        \label{fig:2d_hubb_det}
\end{figure}

We use the Hubbard Models (HMs) as described in Section~\ref{ssec:HM} with varying grid arrangements and fillings; and thus varying numbers of parameters. Depending on the grid  and the filling used, the energy landscapes have different complexities. In Fig.~\ref{fig:2d_hubb_det} we show  approximations of the energy landscape for the simplest case of a 2x1 grid, with a (1,0) and (1,1) filling, respectively, for a  run of the HM without (left images) and with (right images)  noise.  We can see that measurement noise changes the energy landscape and can  make it more difficult to optimize by introducing local optima.  The contours in the figures were created with  the Gaussian process model approximation of the energy landscapes using the combination of squared exponential  and  white noise kernels.

We use six different HM examples of varying dimensionality and optimization difficulty in our numerical study.
Larger grids require more qubits to simulate ($2$ qubits per site in the basic encoding) and will have more optimization parameters, thus allowing us to study the scaling of the proposed methods without fundamentally altering the scientific problem considered.
The number of electrons added to the model (the ``filling") provides some level of control on the total number of optimization parameters: electrons of the same quantum state are indistinguishable and the filling thus determines overall symmetry of the problem.
Higher symmetry means fewer, lower symmetry means more optimization parameters.
We consider two sets of test cases, namely a noise-free test set where the simulated energy is a deterministic output, and a second set where we add measurement noise  $\sigma_s=0.003$, realistic {\em after} unfolding measurement data. 
Table~\ref{tab:testspec} shows the specifics of each test problem we are working with.
Further noise is introduced by restricting each evaluation to 8192 samples (the default on IBM devices).

\begin{table}[htbp]
    \centering
    \begin{tabular}{cccc}
    \hline
       Problem ID  & Grid &Filling & \# Parameters  \\
\hline
H1-d / H1-n  & 2x1 & (1,0) & 2\\
H2-d / H2-n & 2x1 & (1,1) & 2\\
H3-d / H3-n  & 2x2 & (1,1) & 9\\
H4-d / H4-n  & 2x2 & (2,2) & 14\\
H5-d / H5-n  & 2x2 & (3,3) & 9\\
H6-d / H6-n & 3x2 & (1,1) & 20\\
\hline
    \end{tabular}
    \caption{Specifics of the HM     test problems investigated in the numerical study. ``d'' indicates the deterministic case, and ``n'' indicates the noisy case.}
    \label{tab:testspec}
\end{table}

\subsection{Deterministic Hubbard Model}

\begin{figure}[htbp]
         \centering
         \hspace{-0.07in}
         \includegraphics[scale=.108]{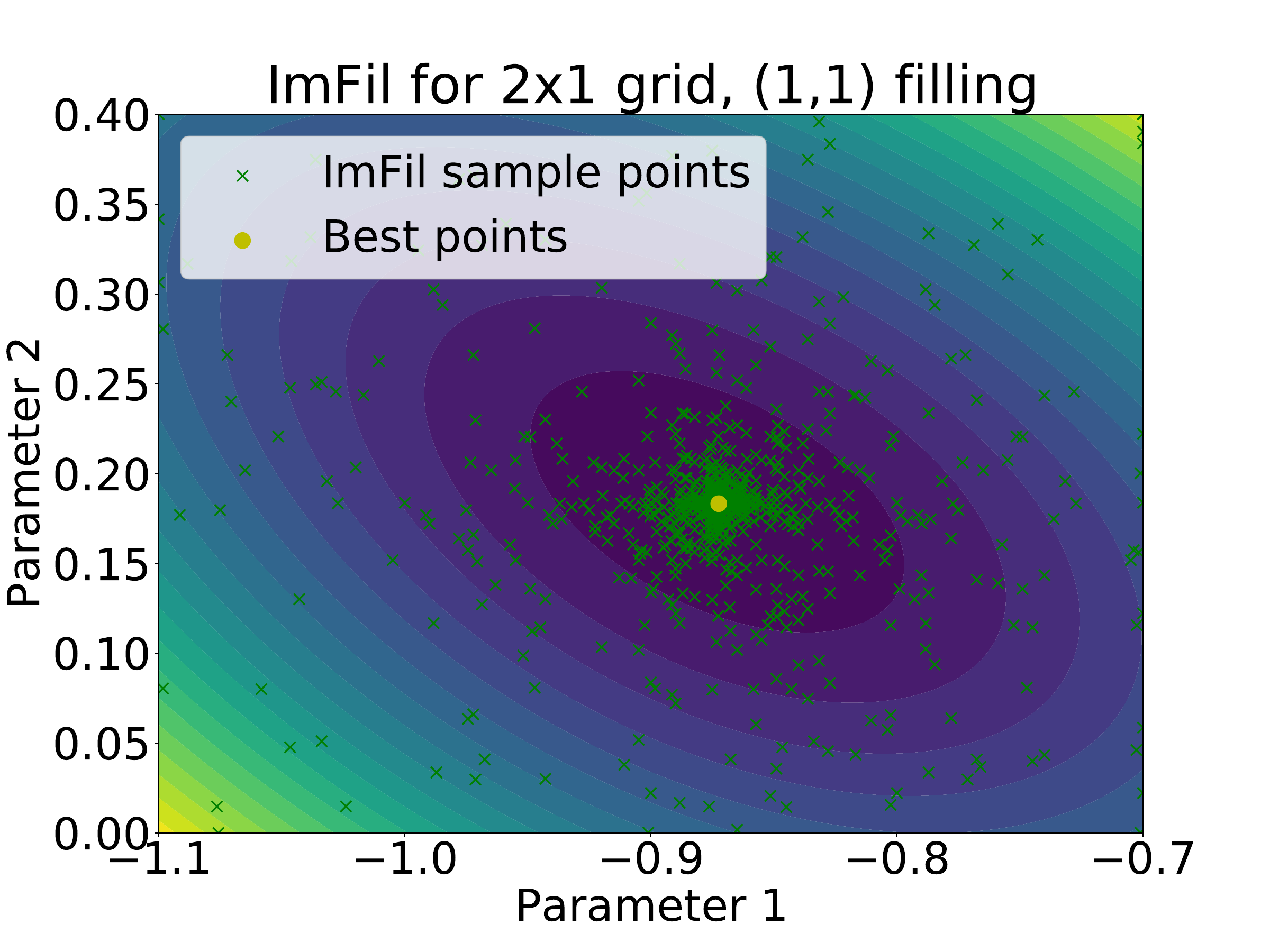}
         \includegraphics[scale=.108]{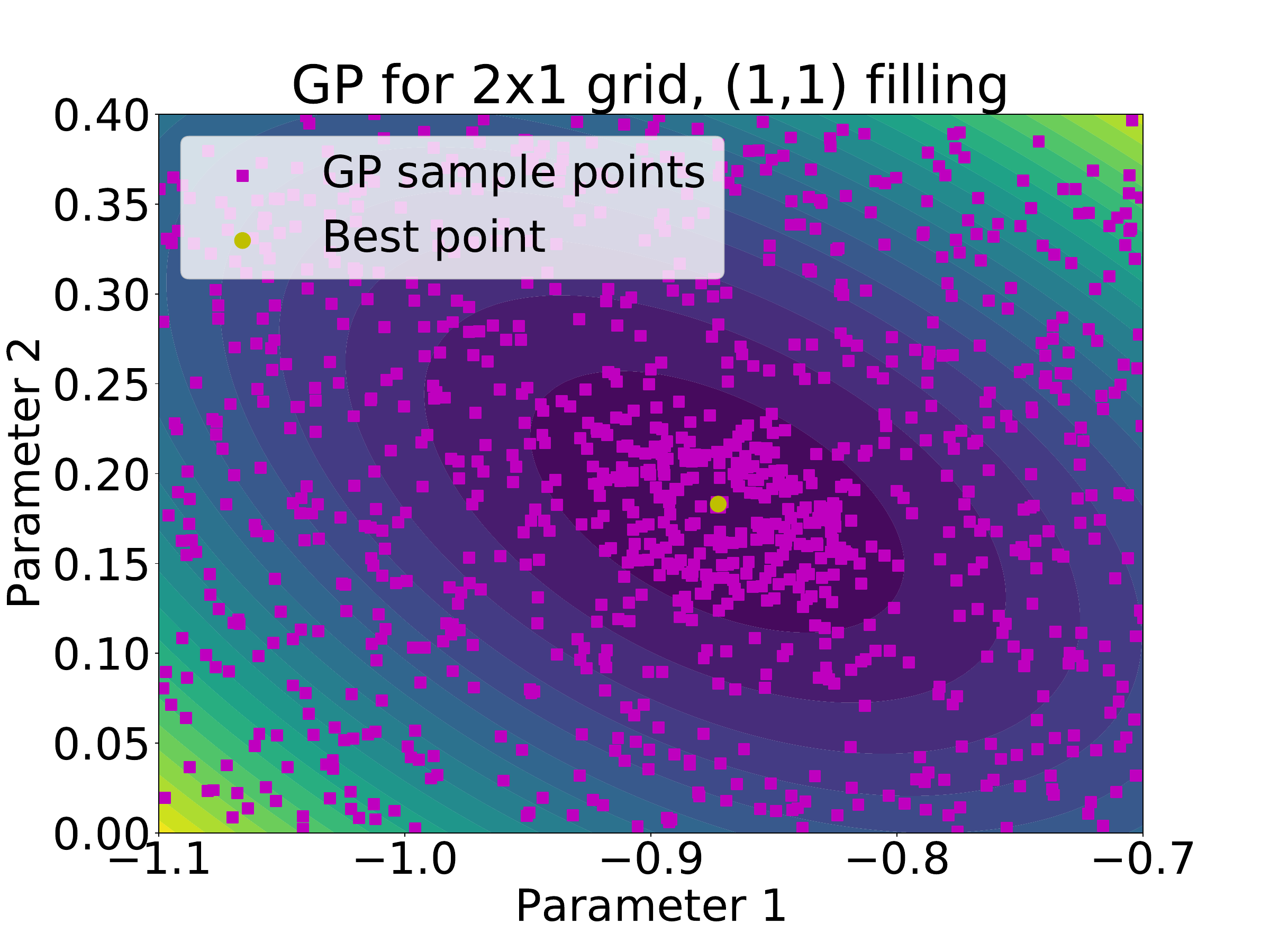}
         \caption{Points sampled by the ImFil method (left) and the Bayesian optimization method using only GP iterations without ImFil search (right), for problem H2-d. The ImFil method samples throughout the whole space, but focuses the search on the location of the minimum. BO trades off exploration and exploitation, thus the samples are more spread out.}
         \label{fig:Imfil_GP_sep_det_2x1_11}
\end{figure}

\begin{figure}[htbp]
         \centering
         \includegraphics[scale=.18]{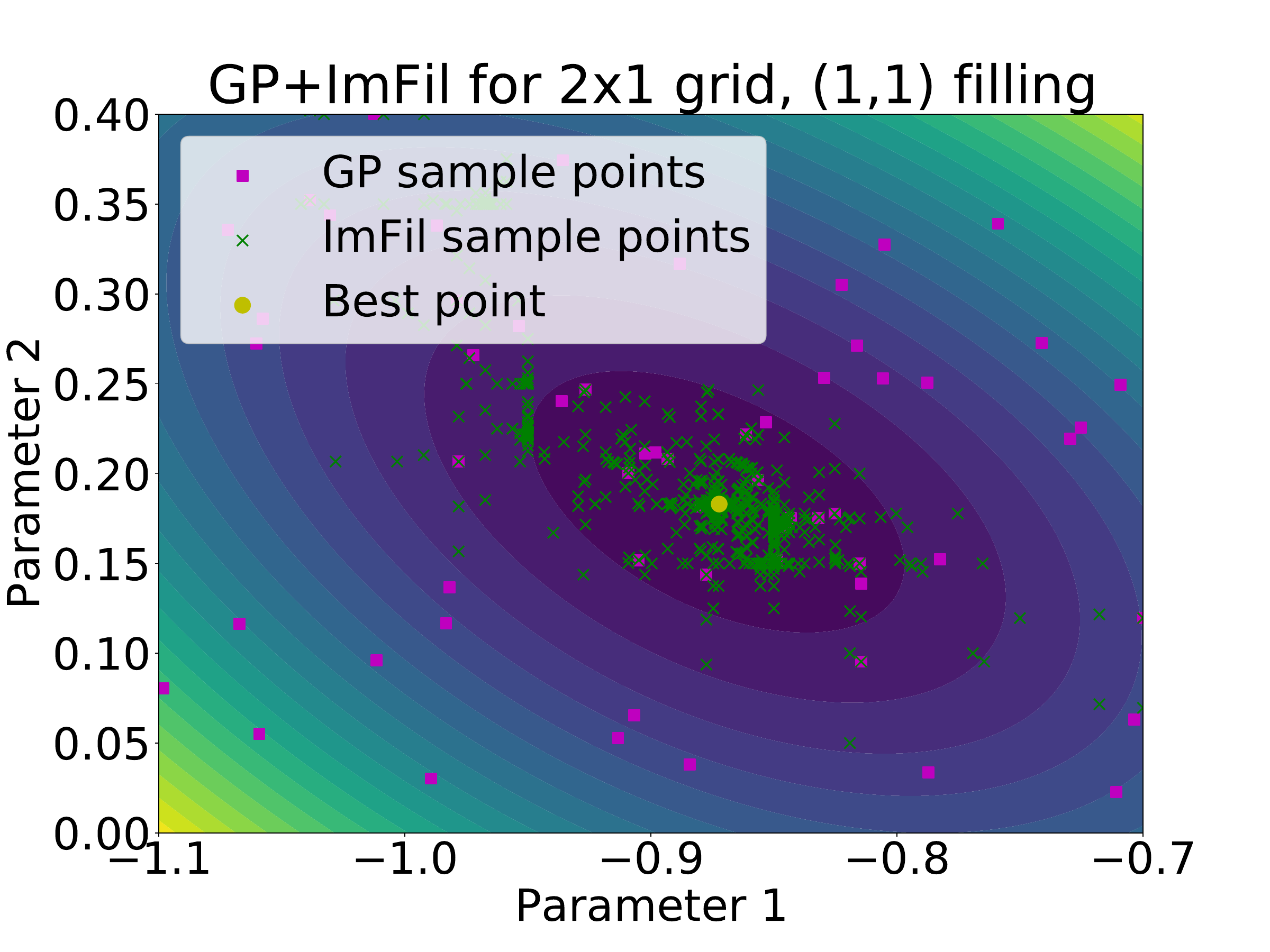}
         \caption{Points sampled by  GP+ImFil  for problem H2-d. The GP points are acquired first, and a subset is used to start the search with ImFil. Each ImFil search is restricted to a smaller hyperbox of the search space to improve the local search behavior.}
         \label{fig:GP_Imfil_det_2x1_11}
\end{figure}

We  ran numerical experiments with the proposed algorithm on the deterministic version of the HM.   
In figures~\ref{fig:Imfil_GP_sep_det_2x1_11} and \ref{fig:GP_Imfil_det_2x1_11}, we show the points sampled by the  three different sampling methods on  the two-dimensional HM example H2-d. The pink square markers in Fig.~\ref{fig:GP_Imfil_det_2x1_11} indicate the samples acquired with the GP iterations.   We can see that the GP samples are well distributed in the space, with denser sampling around the optimum (yellow point), indicating that the GP iterations moved  toward the optimal solution.  We also see multiple clusters of ImFil samples (green crosses), indicating the restarts of the local search   from several GP points that were chosen as outlined in Section~\ref{ssec:globloc}. Note that the green crosses do not cover the full parameter space, but rather they are constrained to small subdomains of the space. In this example, we could have stopped the algorithm after the first local search  concluded (the problem is unimodal and this first search started in the vicinity of the optimum). On the other hand, as shown in Fig.~\ref{fig:Imfil_GP_sep_det_2x1_11} (left), when using ImFil only without the GP iterations or restricting its search to subdomains, the   samples are taken across  the whole space. There is a much larger and denser cloud of points near the optimum (indicated by the yellow point).  When using the Bayesian optimization (Fig.~\ref{fig:Imfil_GP_sep_det_2x1_11}, right),  we see that  samples are collected  all over the space, with  denser sampling in the vicinity of the optimum, but not as dense as for ImFil. Although the Bayesian optimization can guide the search towards the local optimum, it often does not sample densely enough to reach a solution accuracy as high as ImFil. We postulate that this is due to the multi-modality of the expected improvement acquisition function and lower uncertainty estimates in densely sampled regions such as the immediate neighborhood of the best solution found. 

Table~\ref{tab:det_energy}  
shows for each of our deterministic test problems the average and standard deviation of the best energy value found. 
Smaller numbers are better.  
We can see that the  GP iterations prior to starting ImFil (GP+ImFil) did in fact not improve the optimization outcomes for most deterministic problems as compared to the results obtained with ImFil. Only for problem H6-d (the largest dimensional problem), did our proposed method find a better solution.   The small dimensional problems (H1-d and H2-d) may be too simple, and all methods find the optimum. Also, in these examples, we restricted the total size of the search space $\Omega$ such that only one minimum was contained.  

\begin{table}[htbp]
    \centering
\adjustbox{max width=.48\textwidth}{
    \begin{tabular}{c|cc|cc|cc}
    \hline
         & \multicolumn{2}{c}{GP+ImFil} & \multicolumn{2}{c}{ImFil} &\multicolumn{2}{c}{GP}\\
          \hline
           ID & mean & std &mean & std &mean & std\\
          \hline
    H1-d     & \textbf{-1.0} & 0 & \textbf{-1.0} & 0 & \textbf{-1.0}& 0\\
    H2-d     & \textbf{-1.23572} & 0 & \textbf{-1.23572} & 0 & \textbf{-1.23572} & 4.7140e-7\\
    H3-d     &    -3.61760 &0.00021 & \textbf{-3.61789} & 0 & -3.61232  & 0.00222\\
    H4-d     & -2.73595 & 0.00346 & \textbf{-2.74114} & 2.4944e-6 & -2.72717 & 0.00252 \\
    H5-d     &  0.37357 & 0.00052 & \textbf{0.37292} & 2.1602e-6 & 0.38719 & 0.00310\\
    H6-d &  \textbf{-5.73215} & 0.00094 & -5.70090 & 0.04845 & -5.70082 & 0.00498 \\
    \end{tabular}}
    \caption{Best $E$ value found for noise-free HMs. Means and standard deviations are computed over 3 trials.}
    \label{tab:det_energy}
\end{table}

\begin{figure}
    \centering
    \vspace{-0.1in}
    \includegraphics[scale=.19]{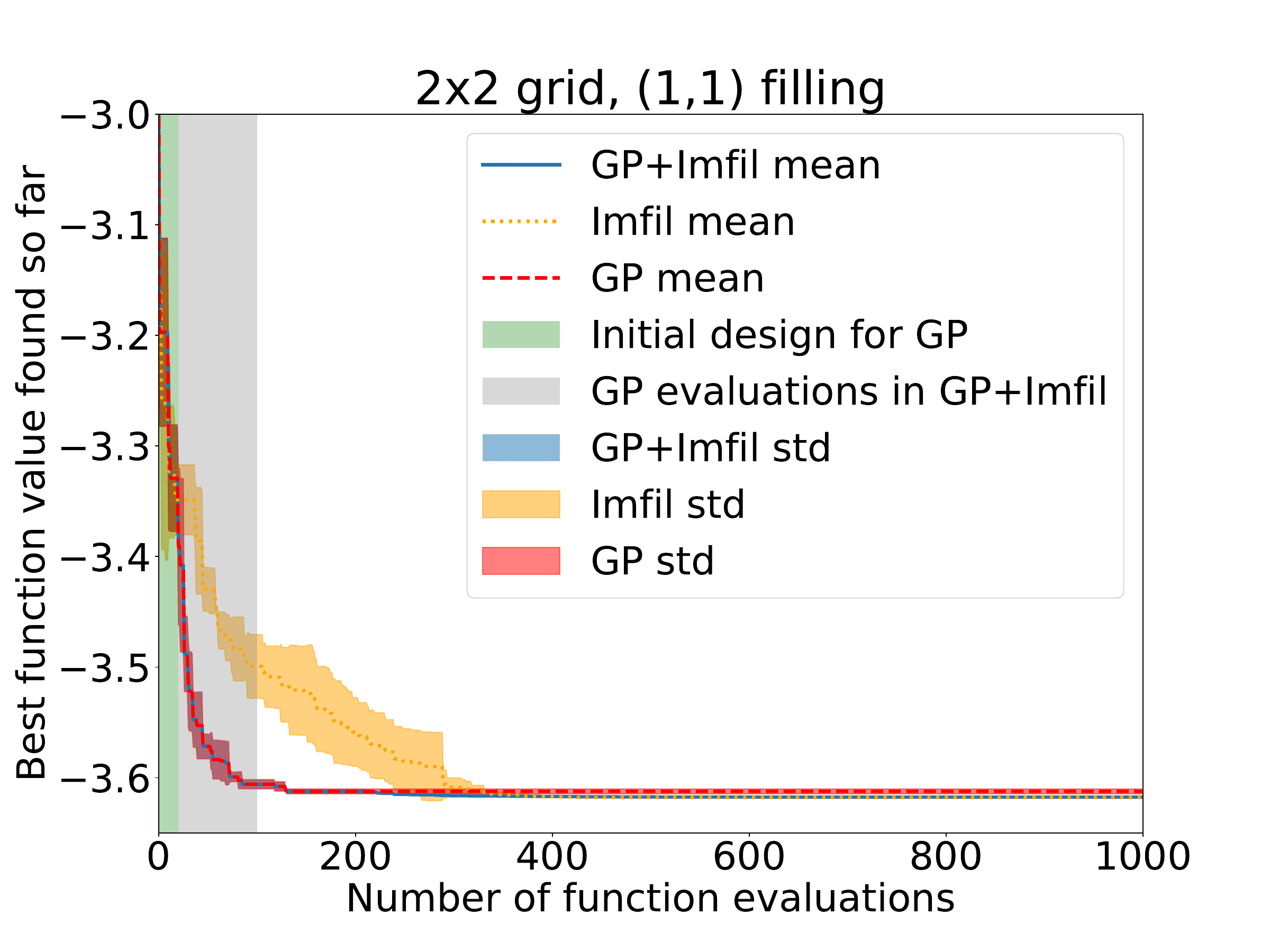}
    \caption{Progress plot for problem H3-d. Lower is better.}
    \label{fig:progress_H3-d}
    \vspace{-0.2in}
\end{figure}
\begin{figure}
    \centering
    \includegraphics[scale=.19]{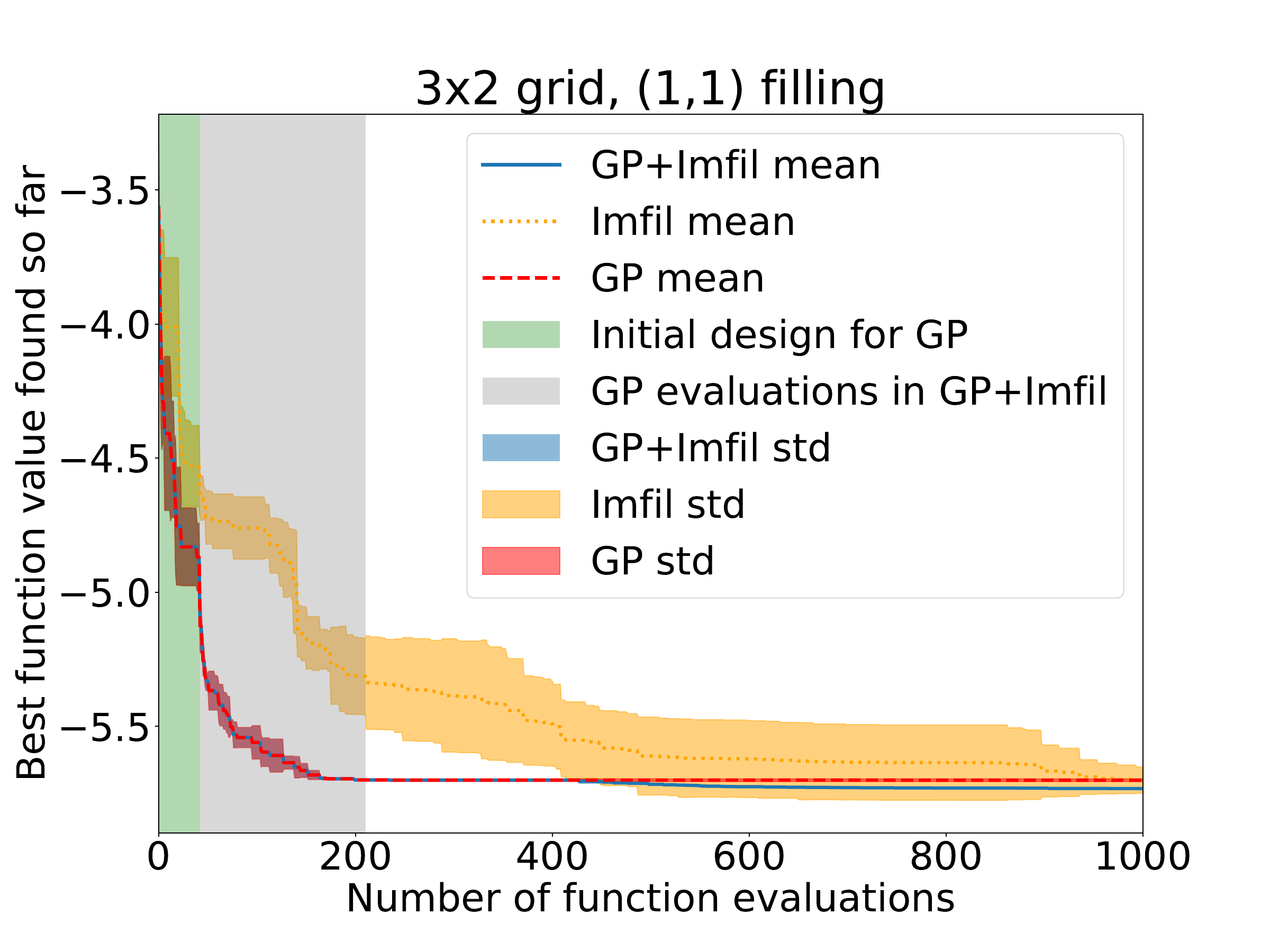}
    \caption{Progress plot for problem H6-d. Lower is better.}
    \label{fig:progress_H6-d}
\end{figure}

Figures~\ref{fig:progress_H3-d} and~\ref{fig:progress_H6-d} show progress plots of the algorithms for two representative problems, namely H3-d and H6-d, respectively. The progress plots show the best function value found so far, and the goal for each graph is to go as low as possible as quickly as possible, which indicates that improvements are found faster. In both figures, we can see that the graphs for  GP and GP+ImFil drop off faster than for ImFil  alone. Both GP+ImFil and GP also have narrower standard deviations than ImFil, which is illustrated with the shaded bands.  However, as can be seen from Table~\ref{tab:det_energy}, ImFil eventually finds for all 2x2-grid problems better final solutions than GP+ImFil. For all problems, using ImFil after the GP iterations led to improvements over the best solution found with the GP iterations. One reason for GP+ImFil not finding as good solutions as ImFil may be related to the local search box being too small and the optimum may be  outside the box. An opportunity to improve this may be the use of adaptive trust regions that can dynamically be expanded, contracted, and moved around depending on the performance of the sample points collected. 




\subsection{Hubbard Model with Measurement Noise}
In our second set of experiments, we consider the same test problems, but we add measurement noise to the energy objective function values.
In the simulation, we apply a misclassification remainder of $0.3\%$, which should be considered as representative after applying standard unfolding techniques~\cite{Unfolding2019}; and we sample $8192$ shots, the default on IBM devices, for each Pauli-string component of the Hamiltonian.

\begin{figure}[hbp]
     \centering
     \vspace{-0.1in}
     \hspace{-0.06in}
    \includegraphics[scale=.108]{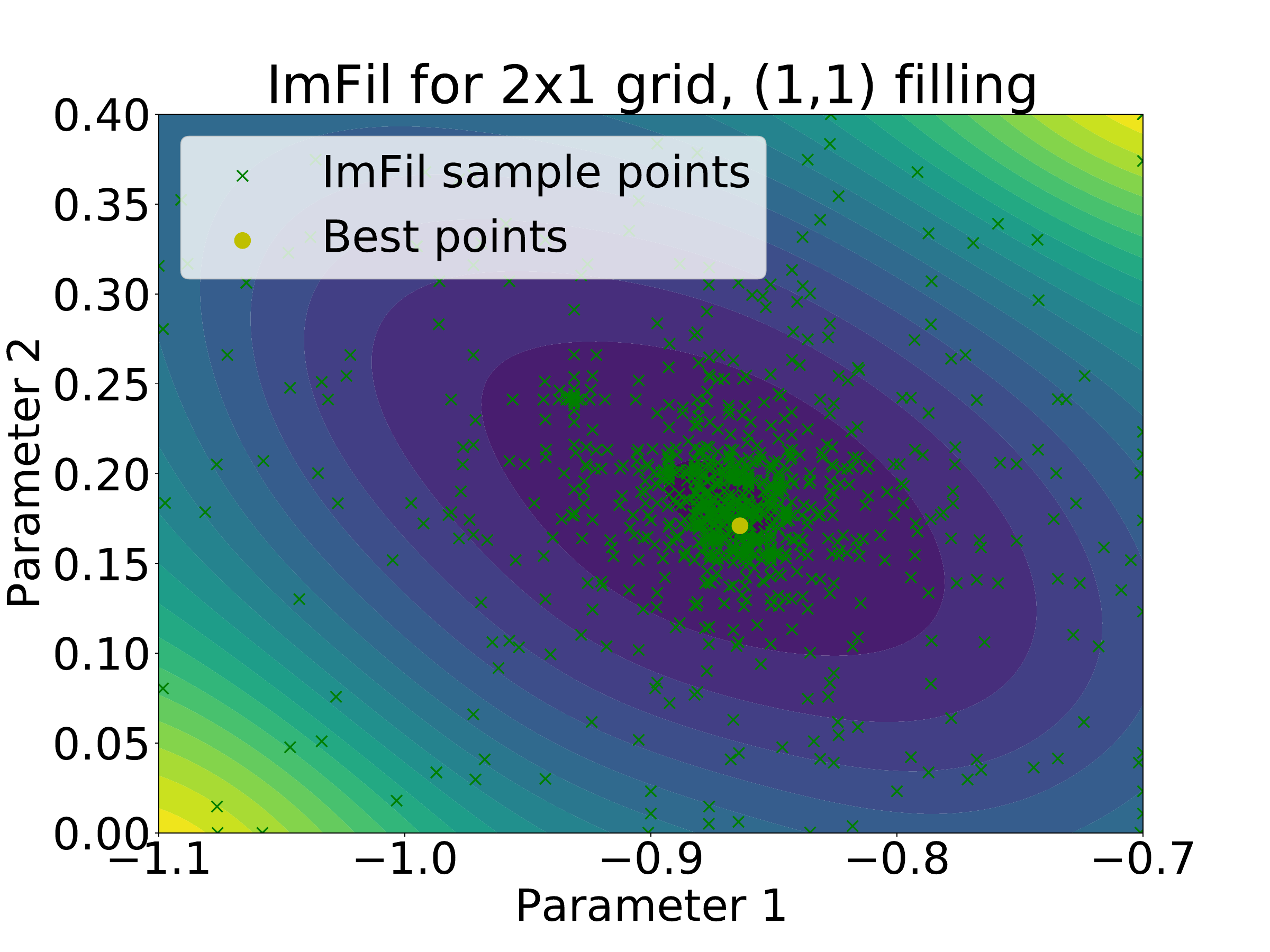}
    \includegraphics[scale=.108]{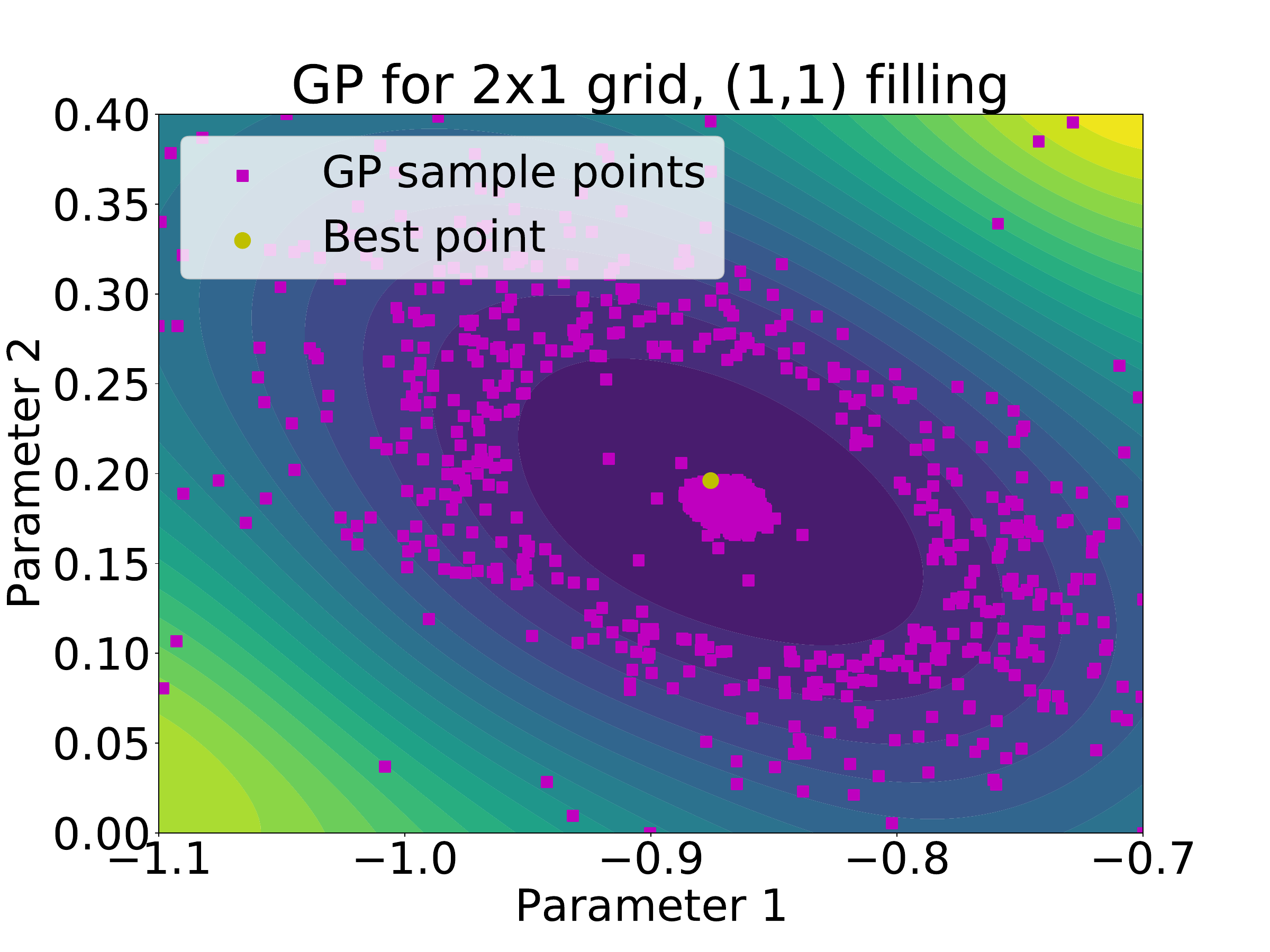}
        \caption{ImFil (left) and GP (right) samples acquired for  problem H2-n.}
        \label{fig:imfil_gp_sep_noise_2x1_11_2}
\end{figure}

\begin{figure}[bp]
     \centering
     \vspace{-0.1in}
    \includegraphics[scale=.19]{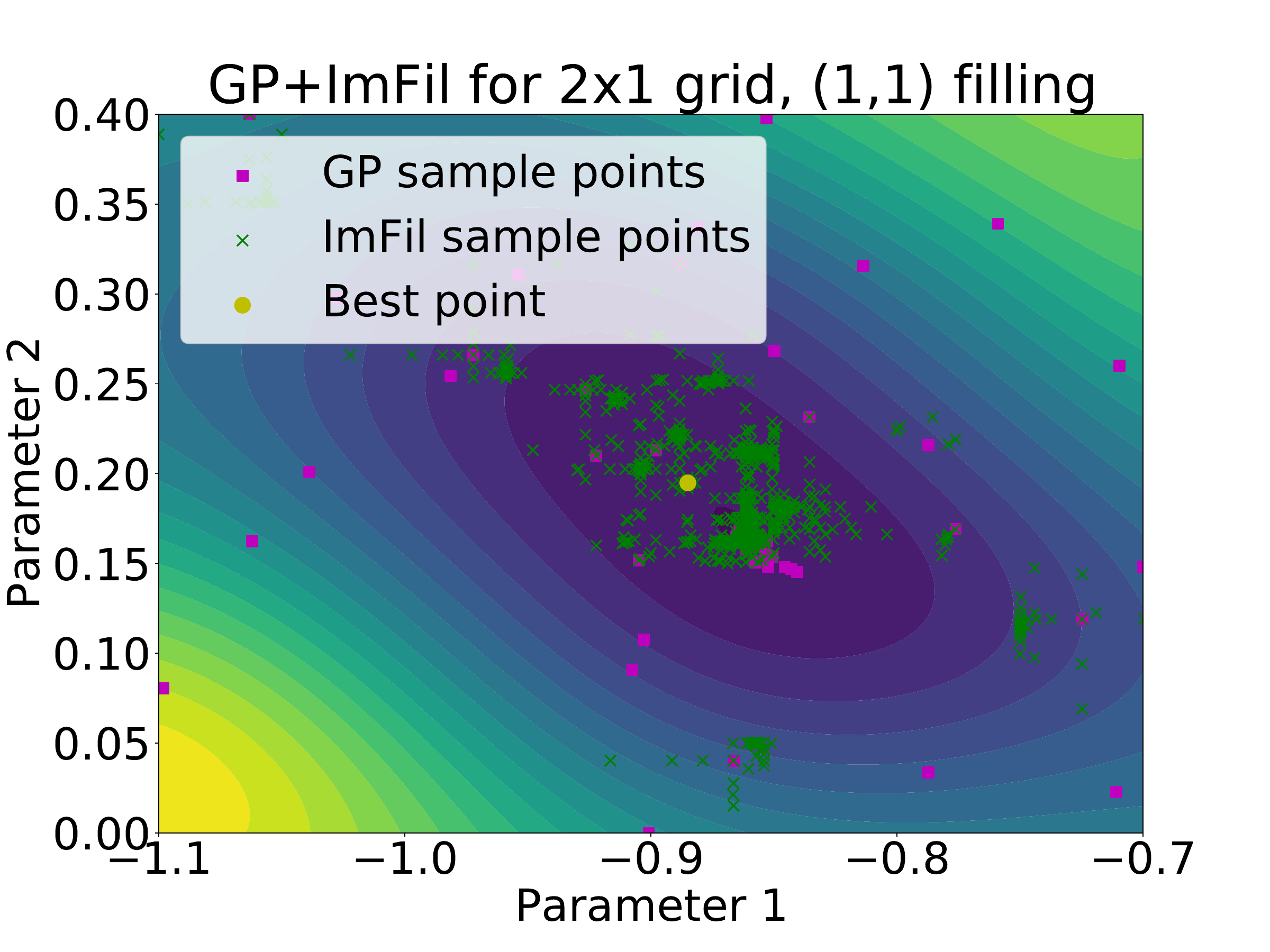}
        \caption{GP+ImFil samples acquired for problem H2-n.}
        \label{fig:gpimfil_noise_2x1_11_2}
\end{figure}

Figures~\ref{fig:imfil_gp_sep_noise_2x1_11_2} and \ref{fig:gpimfil_noise_2x1_11_2} show the samples taken by all three algorithms for the two-dimensional example problem H2-n. In Fig.~\ref{fig:gpimfil_noise_2x1_11_2}, we can see that the GP-based sampling initially explores the full space and then starts multiple local searches with ImFil in the vicinity of the GP's minimum.  In contrast to the deterministic case (Fig.~\ref{fig:GP_Imfil_det_2x1_11}), there appear to be more local search  restarts with ImFil, which is indicated by a larger number of clusters of green crosses in Fig.~\ref{fig:gpimfil_noise_2x1_11_2}.

\begin{figure}[htbp]
     \centering
     \vspace{0.05in}
     \hspace{-0.05in}
    \includegraphics[scale=.26]{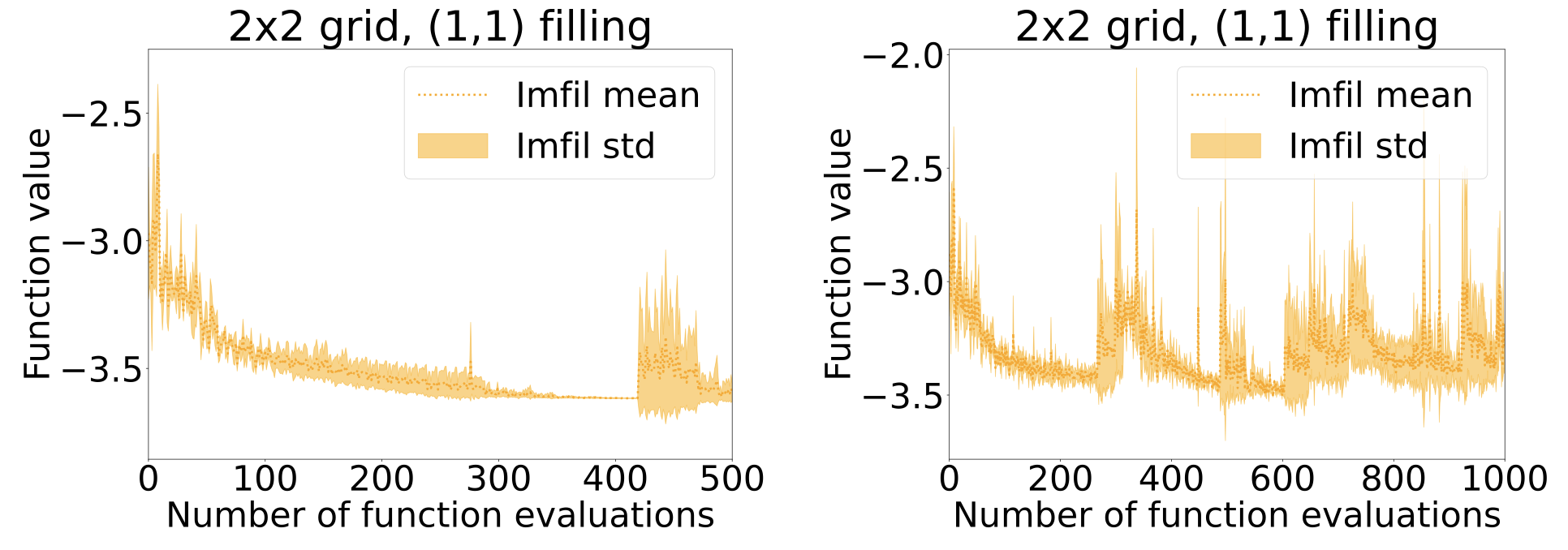}
        \caption{Raw ImFil function values. The deterministic case (left) has fewer restarts of ImFil  than the noisy case (right)  for problem H3-d/n.}
        \label{fig:imfil_unsorted}
    \vspace{-0.1in}
\end{figure}

We can observe a similar behavior in Fig.~\ref{fig:imfil_gp_sep_noise_2x1_11_2} (left), where the samples taken by ImFil are less dense in the vicinity of the optimum than  in Fig.~\ref{fig:Imfil_GP_sep_det_2x1_11} (left) and smaller clusters of points appear more spread out.  This  indicates that for the noisy case, each ImFil run performs fewer function evaluations, and quickly converges to local noise-induced minima. This is illustrated in Fig.~\ref{fig:imfil_unsorted}, where we show the raw ImFil function values (instead of the best function value found so far) for \mbox{H3-d} (left) and H3-n (right). We see clearly that the number of ImFil restarts is lower for the deterministic case (indicated by large jumps in the function values) than for the noisy case.

The samples of the Bayesian optimization (Fig.~\ref{fig:imfil_gp_sep_noise_2x1_11_2}, right) show an interesting behavior where many samples are taken in a circular fashion around the vicinity of the optimum, and the location of the optimum is  densely sampled.

In Table~\ref{tab:noise_energy},  
we show the mean and standard deviation of the best energies found with each algorithm. 
Unlike for the deterministic case, the GP+ImFil method finds the best solution for all higher-dimensional problems (9 and more dimensions). Compared to the deterministic case, the best solutions are also found faster (i.e.\ convergence is reached faster for these problems). 

\begin{table}[htbp]
    \centering
     \adjustbox{max width=.48\textwidth}{
    \begin{tabular}{c|cc|cc|cc}
    \hline
         & \multicolumn{2}{c}{GP+ImFil} & \multicolumn{2}{c}{ImFil} &\multicolumn{2}{c}{GP}\\
          \hline
           ID & mean & std &mean & std &mean & std\\
          \hline
    H1-n     &  -1.01038 & 0.00708 & -1.00907 & 0.00430 & \textbf{-1.01359} & 0.00407 \\
    H2-n     &  -1.24243 & 0.00285  & \textbf{-1.24495} & 0.00404 &  -1.23938&0.00137\\
    H3-n     & \textbf{-3.57227} & 0.00346 & -3.54537 & 0.01703 & -3.54525 & 0.01407\\
    H4-n     & \textbf{-2.70915} & 0.01832 & -2.64677 & 0.06159 & -2.68148 & 0.02624\\
    H5-n     &  \textbf{0.36353} & 0.00728 & 0.38615 & 0.01560 & 0.41325 & 0.00509 \\
    H6-n & \textbf{-5.59416} & 0.01203 & -5.49337 & 0.01221 & -5.55098 & 0.01702 \\
    \end{tabular}}
    \caption{Best energy value $E$ for noisy HM simulations. Mean and standard deviations are computed over 3 runs each.}
    \label{tab:noise_energy}
\end{table}


In figures~\ref{fig:progress_meas_2x2_(1,1).pdf} and~\ref{fig:progress_meas_3x2_(1,1).pdf}, we illustrate the convergence plots of the different algorithms for problems H3-n and  H6-n. Similar to the deterministic case, GP+ImFil and GP find improvements faster than ImFil and ImFil's performance variability is larger.  The Bayesian optimization method gets stuck as is evidenced by the flat line after the first couple of improvements and the addition of multiple ImFil searches improves the performance as is evidenced by the GP+ImFil graphs. From the progress plots of the higher-dimensional problems (figures~\ref{fig:progress_H6-d} and \ref{fig:progress_meas_3x2_(1,1).pdf}), we can see that GP+ImFil does not improve during the second half of the GP iterations. This indicates that we could have potentially found further improvements quicker, had we stopped the GP iterations earlier. Thus, one could experiment with a dynamic stopping criterion that exists the GP iterations after a predefined number of failed iterative improvement trials. However, also this predefined number is a parameter that must be determined. 

\begin{figure}[htbp]
     \centering
    \includegraphics[scale=.19]{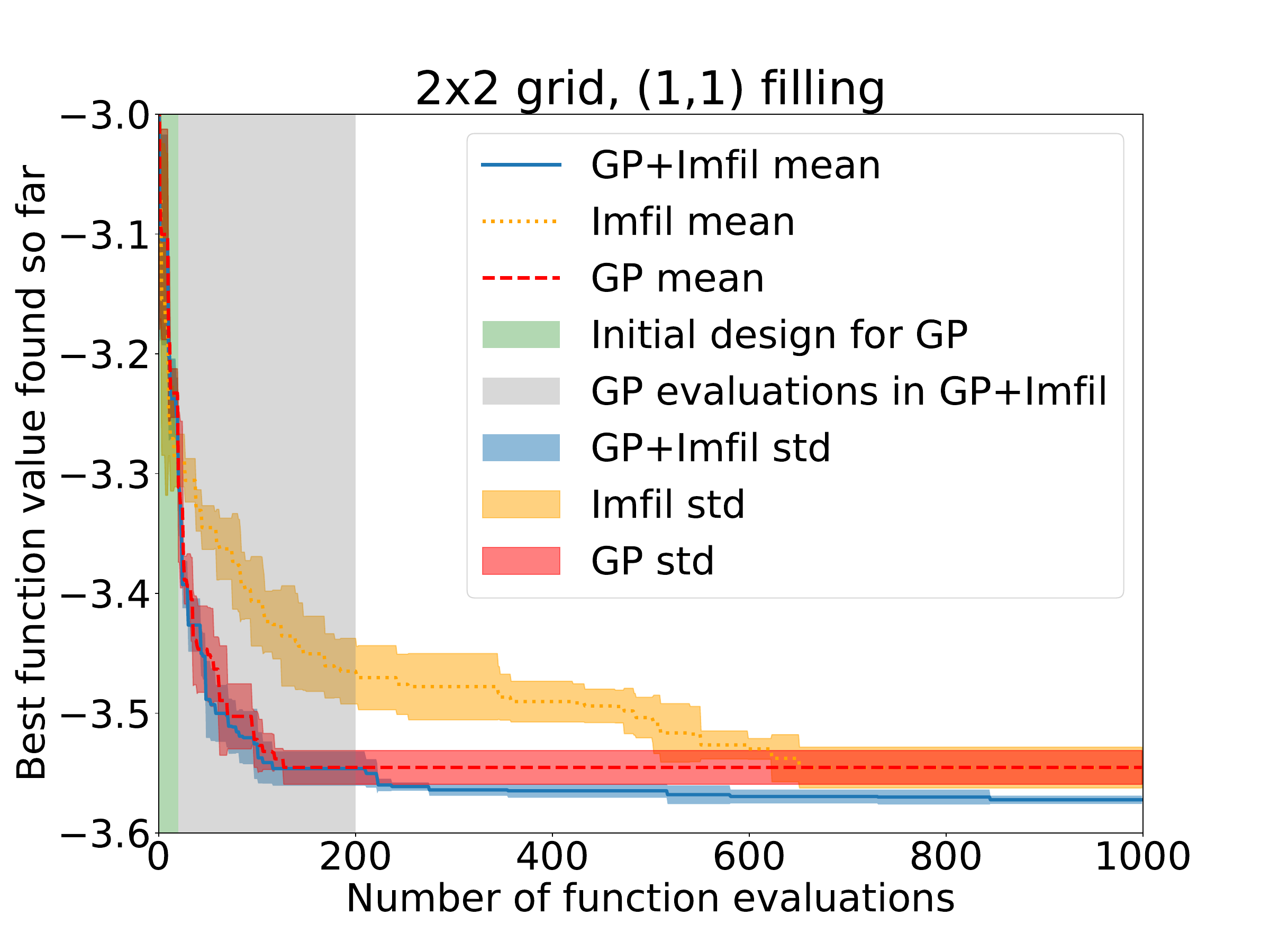}
        \caption{Progress plot for problem H3-n. Lower is better. }
        \label{fig:progress_meas_2x2_(1,1).pdf}
\end{figure}

\begin{figure}[hbp]
     \centering
     \vspace{-0.2in}
    \includegraphics[scale=.19]{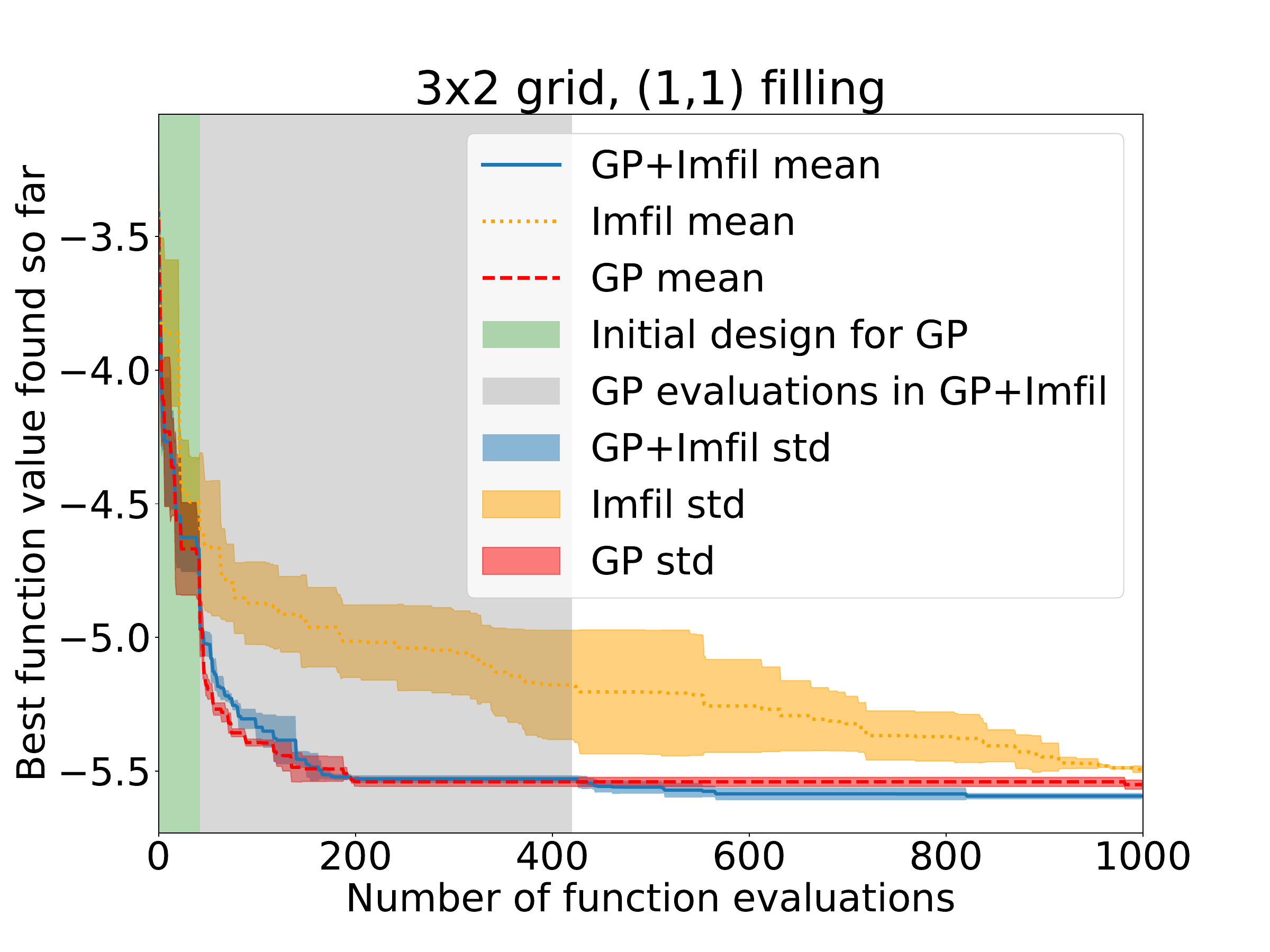}
        \caption{Progress plot for problem H6-n. Lower is better. }
        \label{fig:progress_meas_3x2_(1,1).pdf}
\end{figure}

\subsection{Effectiveness of Seeding}
The GP provides seeds to multistart ImFil, increasing the likelihood to find the true global minimum in a multi-modal landscape.
To study the effect of seeding in isolation, we consider the deterministic case for the 2x2 sized grid configurations, but now using a larger bounding box $\Omega$ spanning the full physical space, which includes many local minima due to  symmetries.
Choosing a larger space requires a different choice of hyperparameters.
In particular, the number of (initial) GP iterations should increase proportionally, to map the larger space to the same level of detail as the smaller problem.
To guarantee that each seeding is run to completion, we do not restrict the local budget $B_{\text{loc}}$, but do limit their number to $5$. 

In Table~\ref{tab:seeding} we compare the results of GP+ImFil v.s.\ ImFil standalone.
In two of the three configurations tested, without a good initial, ImFil got stuck in a local minimum, whereas in all cases GP+ImFil was able to find the true global minimum.
Due to the non-deterministic behavior of GP, however, success is not guaranteed and for two configurations, multiple runs were necessary.
The most common reason for failure is the selection of seeds close to the overall boundary: if there is a downward slope to follow, then the point at the boundary will form an artificial local minimum.
These solutions are easy to flag, however, and the algorithm could reduce the weighting in the seed selection of points originating from extrapolation towards the boundary.
\begin{table}[htbp]
    \centering
    \adjustbox{max width=.48\textwidth}{
    \begin{tabular}{c|c|cc|ccc}
    \hline
      ID & Best on & \multicolumn{2}{c|}{ImFil} &  \multicolumn{3}{c}{GP+ImFil} \\
         & Ansatz  & Result & Iters & Result & Iters & Rate \\
    \hline
    H3-d & -3.62653 & -3.39149 & 375 & -3.62235 & 3407 & 100\% \\
    H4-d & -2.80081 & -2.09527 & 1015 & -2.68146 & 5023 & 20\% \\
    H5-d &  0.37282 &  0.37284  & 535 & 0.37325 & 2480 & 40\% \\
    \end{tabular}}
    \caption{Average results obtained for the deterministic case for a full range global search for successful searches.}
    \label{tab:seeding}
\end{table}

The results shown are an average, but these are not all from the same parameters (there are multiple global minima in the full search space because of symmetry and periodicity).
In practical use, it would make sense to re-run the experiment several times on a single good seed with parameters furthest from the bounds, for the best final result.

A true exhaustive multistart method divides up the full parameter search space and starts searches in each region.
Such an approach does not suffer from a limiting success rate, but at a (much) greater resource cost.
From our results it is clear that the seeding is not as effective as such a multistart would be and  the selection of the initial guesses could be modified to find better performance.
However, since ImFil is restricted at each seed to a boundary box that is only a small subdomain of the original problem (here chosen to be  $\pm 0.2$  in each parameter), even for the lowest observed success rate of 20\%, the resource costs of multiple runs is still vastly lower.\footnote{Since ImFil standalone will stop when converged, if it does not get stuck in a local minimum, it is still the preferred approach.}

For scientific problems where the obtained minimum can not be easily verified, the guarantee of a true multistart will outweigh its resource cost.
However, if such verification is possible, the GP approach outperforms.

\subsection{A Note on Computational Time}
The compute time to acquire points with ImFil and the GP differ significantly. While ImFil uses a simple coordinate search to approximate the gradients, the GP model must be trained on all data (which requires solving a maximum likelihood problem), and, once trained, another optimization problem must be solved to find the next sample point. For the small problems with fast function evaluations, such as H1/2, using a method that involves the GP is therefore not recommended. As we have shown, using the GP for these simple problems does not yield a significantly better performance and thus ImFil may be sufficient as the sampler. On the other hand, for problems where the computational overhead of function evaluations is large, the GP's overhead quickly becomes negligible and the improvements in performance, especially for the noisy cases motivate the use of GP. 

\subsection{Other Kernels}
\begin{figure}[htbp]
    \centering
    \vspace{-0.1in}
    \includegraphics[scale = .2]{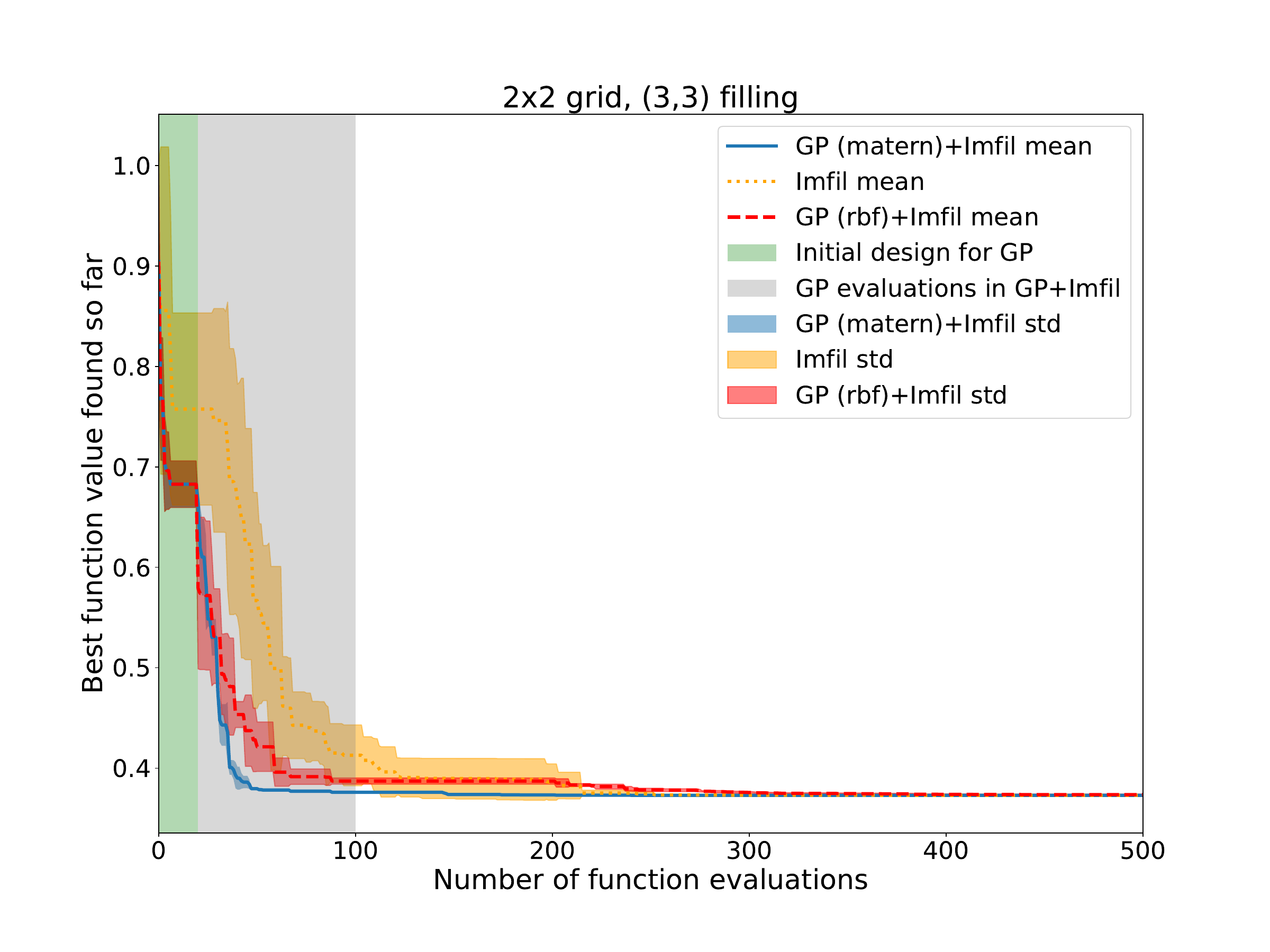}
    \caption{Progress plot for problem H5-d, showing the performance of GP+ImFil with Mat\'ern kernel (blue), RFF+White Kernel (red), and ImFil only (yellow). Lower is better.}
    \vspace{-0.1in}
    \label{fig:matern_prelim}
\end{figure}

In Table~\ref{tab:det_energy_mat} we show initial, exploratory results that we obtained for the 2x2 deterministic test cases when using the Mat\'ern kernel in the GP+ImFil method (column 2).
This kernel type can be interpreted as a generalization of the RBF kernel, and can better capture physical processes due to its finite differentiability for a range of finite parameter settings.
We can see that with the Mat\'ern kernel, the results are better than when using the RBF+White Kernel (column 4). In Fig.~\ref{fig:matern_prelim}, we show a convergence graph for problem 2x2 with (3,3) filling. We can see that convergence is quicker than for  ImFil. For the noisy version of the problem, the Mat\'ern kernel did not yield improvements. However, further investigation into this kernel choice is needed for definite conclusions. 

\begin{table}[htbp]
    \centering
\adjustbox{max width=.48\textwidth}{
    \begin{tabular}{c|cc|cc|cc}
    \hline
         & \multicolumn{2}{c}{GP (Mat\'ern)+ImFil} & \multicolumn{2}{c}{ImFil} &\multicolumn{2}{c}{GP (RBF+WK)+ImFil}\\
          \hline
           ID & mean & std &mean & std &mean & std\\
          \hline
    H3-d     &    \textbf{-3.61789} &8.165e-07 & \textbf{-3.61789} & 0 &  -3.61760 &0.00021 \\
    H4-d     & -2.74113 & 1.203e-05 & \textbf{-2.74114} & 2.4944e-6 & -2.73595 & 0.00346 \\
    H5-d     &  \textbf{0.37292} & 8.165e-07& \textbf{0.37292} & 2.1602e-6 & 0.37357& 0.00094\\
    \end{tabular}}
    \caption{Best $E$ value for noise-free HMs with Mat\'ern kernel, averaged results from 3 trials.}
    \label{tab:det_energy_mat}
\end{table}

\section{Discussion, Conclusions, and Future Directions}\label{sec:concl}
We presented an optimization routine that couples a Gaussian process model based global search with multiple local Implicit Filtering searches (``GP+ImFil'') to solve the  VQE problem for the Hubbard Model. We examined two classes of problems, namely noise-free models and models with measurement and sampling noise, over a range of 2-20 optimization parameters. Our goal was to examine if GP models with appropriately chosen kernels to take into account the noise can help us find optimal solutions with fewer objective evaluations than  widely-used classical optimizers such as ImFil.  

Our results for the deterministic problems showed that in most cases, ImFil finds better solutions than the GP+ImFil approach even though its progress toward the optimal solutions is significantly slower. On the other hand, for the noisy problems, GP+ImFil not only finds improvements faster but also finds overall better solutions than ImFil. For the noisy functions, ImFil does restart more often than in the deterministic case, which may be due to ImFil converging quickly to noise-induced local optima, and therefore a careful selection of starting guesses as done in the GP+ImFil method is beneficial.  Our results indicate that for smooth unimodal surfaces  such as  the deterministic problems with small bounding boxes, the GP iterations are not advantageous. However, for very rugged surfaces such as in the noisy case, using the smooth GP to guide the local search helps us avoid converging to noise-induced local minima that are far from the global optimum. 

There are several potential improvements of our proposed method that are worthwhile to study in the future. First, the GP+ImFil method has various parameters such as the number of GP iterations, the start point selection for ImFil, and the definition of the local search box for ImFil. A more dynamic adjustment of these parameters in direct response to the observed function values may lead to better optimization performance.
In particular, for search bounds, artificial minima should be detected and stopped out; and too restrictive bounds should be dynamically re-adjusted.
Second, one could experiment with other local search methods that were developed  for noisy problems such as SnobFit. SnobFit requires multiple starting points as input and our GP iterations allow us to select these points. 
Third, parallelism must be exploited where possible in order to solve higher dimensional problems. This may go hand in hand with exploiting tools such as GPyTorch~\cite{gpytorch} that allow for faster training of Gaussian process models. 
Finally, we observe that the choice and configuration of the GP kernel matters. Thus, future work should continue the exploration of different types of kernels, such as the Mat\'ern kernel.




\bibliographystyle{plain}
\bibliography{main}

\begin{thebibliography}{10}

\bibitem{Qiskit}
{Abraham, H. {\em et.~al.}}
\newblock Qiskit: An open-source framework for quantum computing, 2021.

\bibitem{bqskit_web}
Bqskit documentation.
\newblock \url{https://bqskit.lbl.gov/}.

\bibitem{Bravyi_2002}
Sergey~B. Bravyi and Alexei~Yu. Kitaev.
\newblock Fermionic quantum computation.
\newblock {\em Annals of Physics}, 298(1):210--226, May 2002.

\bibitem{Cerezo2021}
M.~Cerezo, Andrew Arrasmith, Ryan Babbush, Simon~C. Benjamin, Suguru Endo,
  Keisuke Fujii, Jarrod~R. McClean, Kosuke Mitarai, Xiao Yuan, Lukasz Cincio,
  and Patrick~J. Coles.
\newblock Variational quantum algorithms.
\newblock {\em Nature Reviews Physics}, 3(9):625--644, Aug 2021.

\bibitem{farhi2014quantum}
Edward Farhi, Jeffrey Goldstone, and Sam Gutmann.
\newblock A quantum approximate optimization algorithm, 2014.

\bibitem{Forrester}
A.I.J. Forrester, A.~S\'obester, and A.J. Keane.
\newblock {\em Engineering Design via Surrogate Modelling: A Practical Guid}.
\newblock John Wiley and Sons, Ltd., 2008.

\bibitem{gpytorch}
Jacob~R. Gardner, Geoff Pleiss, David Bindel, Kilian~Q. Weinberger, and
  Andrew~Gordon Wilson.
\newblock {GPyTorch}: Blackbox matrix-matrix gaussian process inference with
  gpu acceleration.
\newblock In {\em Proceedings of the 32nd International Conference on Neural
  Information Processing Systems}, NIPS'18, page 7587–7597, Red Hook, NY,
  USA, 2018. Curran Associates Inc.

\bibitem{Hubbard1963}
J.~Hubbard.
\newblock Electron correlations in narrow energy bands.
\newblock {\em Proc. R. Soc. Lond.}, 276(1365):238–257, Nov 1963.

\bibitem{snobfit}
Waltraud Huyer and Arnold Neumaier.
\newblock Snobfit -- stable noisy optimization by branch and fit.
\newblock {\em ACM Trans. Math. Softw.}, 35(2), Jul 2008.

\bibitem{Iannelli}
Giovanni Iannelli and Karl Jansen.
\newblock Noisy bayesian optimization for variational quantum eigensolvers,
  2021.

\bibitem{Jones2001}
D.R. Jones.
\newblock A taxonomy of global optimization methods based on response surfaces.
\newblock {\em Journal of Global Optimization}, 21:345--383, 2001.

\bibitem{Jones1998}
D.R. Jones, M.~Schonlau, and W.J. Welch.
\newblock Efficient global optimization of expensive black-box functions.
\newblock {\em Journal of Global Optimization}, 13:455--492, 1998.

\bibitem{qiskitSPSA}
Abhinav Kandala, Antonio Mezzacapo, Kristan Temme, Maika Takita, Markus Brink,
  Jerry~M. Chow, and Jay~M. Gambetta.
\newblock Hardware-efficient variational quantum eigensolver for small
  molecules and quantum magnets.
\newblock {\em Nature}, 549(7671):242--246, 2017.

\bibitem{ImFil}
C.T. Kelley.
\newblock {\em Implicit Filtering}.
\newblock SIAM, 2011.

\bibitem{Kubler}
Jonas~M. Kübler, Andrew Arrasmith, Lukasz Cincio, and Patrick~J. Coles.
\newblock An adaptive optimizer for measurement-frugal variational algorithms.
\newblock {\em Quantum}, 4:263, May 2020.

\bibitem{tutorial}
W.~Lavrijsen, J.~M\"uller, and E.~Younis.
\newblock Tutotial: Workflow for hybrid quantum-classical algorithm,
  https://github.com/scikit-quant/scikit-quant/tree/master/tutorials.
\newblock In {\em IEEE International Conference on Quantum Computing and
  Engineering QCE21}, 2021.

\bibitem{Wim}
Wim Lavrijsen, Ana Tudor, Juliane Müller, Costin Iancu, and Wibe de~Jong.
\newblock Classical optimizers for noisy intermediate-scale quantum devices.
\newblock In {\em 2020 IEEE International Conference on Quantum Computing and
  Engineering (QCE)}, pages 267--277, 2020.

\bibitem{Nomad}
S\'{e}bastien Le~Digabel.
\newblock Algorithm 909: Nomad: Nonlinear optimization with the mads algorithm.
\newblock {\em ACM Trans. Math. Softw.}, 37(4), Feb 2011.

\bibitem{lbfgs}
Dong~C. Liu and Jorge Nocedal.
\newblock On the limited memory {BFGS} method for large scale optimization.
\newblock {\em Mathematical Programming}, 45(1-3):503--528, 1989.

\bibitem{Matheron1963}
G.~Matheron.
\newblock {P}rinciples of geostatistics.
\newblock {\em Economic Geology}, 58:1246--1266, 1963.

\bibitem{McClean2015}
Jarrod~R McClean, Jonathan Romero, Ryan Babbush, and Al{\'{a}}n Aspuru-Guzik.
\newblock {The theory of variational hybrid quantum-classical algorithms}.
\newblock {\em New Journal of Physics}, 18(2):23023, 2016.

\bibitem{Mockus}
Jonas Mockus.
\newblock Application of bayesian approach to numerical methods of global and
  stochastic optimization.
\newblock {\em Journal of Global Optimization}, 4(4):347--365, 1994.

\bibitem{mopaper}
J.~M\"uller.
\newblock {SOCEMO: surrogate optimization of computationally expensive
  multiobjective problems}.
\newblock {\em INFORMS Journal on Computing}, 29(4):581--596, 2017.

\bibitem{somi}
J.~M\"uller, C.A. Shoemaker, and R.~Pich\'e.
\newblock {SO-MI: A surrogate model algorithm for computationally expensive
  nonlinear mixed-integer black-box global optimization problems}.
\newblock {\em Computers \& Operations Research, 2013}, 40(5):1383--1400, 2013.

\bibitem{Unfolding2019}
Benjamin Nachman, Miroslav Urbanek, Wibe~A. de~Jong, and Christian~W. Bauer.
\newblock Unfolding quantum computer readout noise, 2019.

\bibitem{Nakanishi}
Ken~M. Nakanishi, Keisuke Fujii, and Synge Todo.
\newblock Sequential minimal optimization for quantum-classical hybrid
  algorithms.
\newblock {\em Physical Review Research}, 2(4), Oct 2020.

\bibitem{nielsen00}
Michael~A. Nielsen and Isaac~L. Chuang.
\newblock {\em Quantum Computation and Quantum Information}.
\newblock Cambridge University Press, 2000.

\bibitem{Peruzzo2013}
Alberto Peruzzo, Jarrod McClean, Peter Shadbolt, Man-Hong Yung, Xiao-Qi Zhou,
  Peter~J. Love, Al{\'{a}}n Aspuru-Guzik, and Jeremy~L. O’Brien.
\newblock {A variational eigenvalue solver on a photonic quantum processor}.
\newblock {\em Nature Communications}, 5(4213):1--7, 2014.

\bibitem{Powell1992}
M.J.D. Powell.
\newblock {\em Advances in Numerical Analysis, vol. 2: wavelets, subdivision
  algorithms and radial basis functions. Oxford University Press, Oxford, pp.
  105-210}, chapter The Theory of Radial Basis Function Approximation in 1990.
\newblock Oxford University Press, London, 1992.

\bibitem{bobyqa}
M.J.D. Powell.
\newblock The {BOBYQA} algorithm for bound constrained optimization without
  derivatives.
\newblock Technical Report DAMTP 2009/NA06, Cambridge University, 2009.

\bibitem{scipy_web}
Scipy documentation.
\newblock \url{https://www.scipy.org}.

\bibitem{shaffer2022}
Ryan Shaffer, Lucas Kocia, and Mohan Sarovar.
\newblock Surrogate-based optimization for variational quantum algorithms,
  2022.

\bibitem{skquant_web}
scikit-quant documentation.
\newblock \url{https://scikit-quant.readthedocs.io}.

\bibitem{SPSA}
J.C. Spall.
\newblock Multivariate stochastic approximation using a simultaneous
  perturbation gradient approximation.
\newblock {\em IEEE Transactions on Automatic Control}, 37(3):332--341, 1992.

\bibitem{Sung_2020}
Kevin~J Sung, Jiahao Yao, Matthew~P Harrigan, Nicholas~C Rubin, Zhang Jiang,
  Lin Lin, Ryan Babbush, and Jarrod~R McClean.
\newblock Using models to improve optimizers for variational quantum
  algorithms.
\newblock {\em Quantum Science and Technology}, 5(4):044008, Sep 2020.

\bibitem{shiro}
Shiro Tamiya and Hayata Yamasaki.
\newblock Stochastic gradient line bayesian optimization: Reducing measurement
  shots in optimizing parameterized quantum circuits, 2021.

\bibitem{Tilly2021}
Jules Tilly, Hongxiang Chen, Shuxiang Cao, Dario Picozzi, Kanav Setia, Ying Li,
  Edward Grant, Leonard Wossnig, Ivan Rungger, George~H. Booth, and Jonathan
  Tennyson.
\newblock The variational quantum eigensolver: a review of methods and best
  practices, 2021.

\bibitem{Learning2Learn}
Guillaume Verdon, Michael Broughton, Jarrod~R. McClean, Kevin~J. Sung, Ryan
  Babbush, Zhang Jiang, Hartmut Neven, and Masoud Mohseni.
\newblock Learning to learn with quantum neural networks via classical neural
  networks, 2019.

\bibitem{Wang_2018}
Zhihui Wang, Stuart Hadfield, Zhang Jiang, and Eleanor~G. Rieffel.
\newblock Quantum approximate optimization algorithm for maxcut: A fermionic
  view.
\newblock {\em Physical Review A}, 97(2), Feb 2018.

\bibitem{Ye2000}
K.Q. Ye, W.~Li, and A.~Sudjianto.
\newblock Algorithmic construction of optimal symmetric {L}atin hypercube
  designs.
\newblock {\em Journal of Statistical Planning and Inference}, 90:145--159,
  2000.

\end{thebibliography}
\end{document}